\documentclass[twocolumn,english,aps,superscriptaddress, pra]{revtex4-1}

\usepackage{array}
\usepackage[latin9]{inputenc}
\usepackage{amsmath}
\usepackage{amssymb}
\usepackage{graphicx}
\usepackage{babel}
\usepackage{mathrsfs}
\usepackage{amsfonts}
\usepackage{epstopdf}
\usepackage{multirow}
\usepackage{color}
\usepackage{natbib}

\begin{document}

\title{Robust Cusps near Topological Phase Transitions:\\
Signatures of Majorana fermions and interactions with fluctuations}

\author{Fan Yang}

\affiliation{Department of Physics and Astronomy, University of British Columbia, Vancouver, British Columbia, V6T 1Z1, Canada}

\author{Shao-Jian Jiang}

\affiliation{State Key Laboratory of Magnetic Resonance and Atomic and Molecular Physics, Wuhan Institute of Physics and Mathematics, Chinese Academy of Sciences, Wuhan, Hubei, 430072, China}

\author{Fei Zhou}

\affiliation{Department of Physics and Astronomy, University of British Columbia, Vancouver, British Columbia, V6T 1Z1, Canada}

\date{\today}

\begin{abstract} 
In this article, we study non-analyticities or cusps near a topological phase transition driven by changes of global topologies rather than by the formation of conventional (local) order.
The particular phase transitions between topologically ordered superfluids and non-topological superfluids studied in this article are always accompanied by the emergence or disappearance of Majorana boundary modes, while at the critical point these boundary modes are liberated and evolve into the bulk.
Proliferation of massless Majorana fermions in the bulk at the transition leaves unique thermodynamic signatures. For a very broad class of isotropic superfluids or superconductors (with or without time reversal symmetry), the resultant transitions in $d$ spatial dimensions will be of $(d+1)$th order. The $(d+1)$th order derivatives of the grand potential are either discontinuous in even dimensions or logarithmically divergent in odd dimensions. 
Bulk quantities such as compressibility can develop a cusp along with the change of topology of the superfluids.
We compute quantitatively the cusp structures and further investigate the effect of interactions induced by massive fluctuation fields.
These cusps are robust as direct manifestations of Majorana fermions and these interactions, and can be potentially considered as an alternative approach for detecting these puzzling particles.

\end{abstract}

\maketitle

\section{Introduction}

Topological quantum matter has been a focus of contemporary studies in condensed matter physics for decades.
It has become well known that, generally speaking, there are two big classes of topological states. One is purely driven by non-trivial twist band structures which can be conveniently characterized by topological invariants such as Chern numbers and $Z_2$ indices. Examples of this class include the quantum Hall effect \cite{Klitzing80, Thouless82}, which explicitly breaks time reversal symmetry, and more recent time reversal symmetric topological insulators which have been extensively studied for more than a decade \cite{Kane05,Bernevig06,Fu07,Moore07,Hasan10,Qi11}. 
The other class of topological states are exclusively driven by interactions, and the topological orders in these systems are solely induced by strong interactions. Fractional quantum Hall effect and spin liquids \cite{Anderson87,Kalmeyer87,Roksar88,Haldane88,Wen89,Moessner01,Read91,Kitaev06a,Balents10,Savary17} are great examples of this class of phenomena. 
The unique aspect of these topological matters is that although some of them can be characterized by non-trivial topological invariants, there can also be an additional topological ground state degeneracy when they live on a torus or other more complicated non-simply connected geometry. 
Such topological degeneracies are also intimately related to the topological entropy evaluated in terms of von Neumann entropy which reflects very intimate connections between the global topology of states and fundamental quantum entanglement \cite{Kitaev06b,Levin06}.

Among the latter class of interacting topological matter, superfluids of $p_x\pm ip_y$ type stands out  as one of the simplest examples\cite{Leggett75}. Like other conventional superconductors and superfluids, it breaks $U(1)$ symmetry spontaneously.
This is unlike many other topological states, interacting or non-interacting, which have non-trivial global topologies but do not support simple local order parameters. Hence, a topological superfluid can have a transition into another superfluid state that breaks the same symmetries but with a different topology. Such an interesting possibility was suggested long ago by the authors of Ref. \cite{Read00,Volovik88}.
 A quantum phase transition like this, which is not driven by simple local ordering, is clearly beyond the standard Landau paradigm of order-disorder phase transitions \cite{Landau,Sachdev}. 
 Although the physical consequence of topological ordering, i.e. the presence of zero energy Majorana edge modes, was pointed out before and became widely known \cite{Read00,Kitaev01,Volovik03,Kopnin91,Volovik99}, what happens to the energetics of the bulk superfluids on the two sides of topological phase transitions seems to have received relatively little attention. 
 As the direct detection of individual Majorana fermions can be quite challenging \cite{Alicea12,Beenakker13,Elliot15}, since they are charge neutral and usually manifest in very peculiar ways, the collective signatures of these fermions in thermodynamics can be an alternatively way to visualize these emergent fermions in nature.

A full  theoretical description of the energetics near such a topological phase transition, and its connections to topology, are what we intend to present in this article.
Our study is mainly based on an infrared effective theory of Nambu (or Majorana) fermions interacting with a fluctuation field in the vicinity of the phase transition. 
An important difference when compared with the standard Wilsonian theory for order-disorder phase transitions is that in our case the $U(1)$ gauge symmetry is broken at the critical point when the topological phase transition occurs between two states that spontaneously break the same symmetries. 
As a result, the effective theory has additional symmetry breaking terms which are usually forbidden in the standard Landau paradigm. 
The detailed contributions of those interactions to topological phase transitions are classified and studied quantitatively in this article.

One of our objectives in this study is to identify a relation between
interactions with fluctuations in terms of Nambu fermions and interactions appearing in terms of Majorana fermions in an effective description. 
The former interactions can be more directly related to how particles interact microscopically. 
We also intend to examine the similarity and/or differences between the interacting Majorana models here and
the ones studied in the traditional field theory context.
For this purpose, we further formulate our effective theory in an irreducible representation involving Majorana fermions interacting with fluctuations of a real scalar field via Yukawa interactions. 
However, as the $U(1)$ gauge symmetry is broken at the critical point, compared to the standard model for interacting Majorana models, our effective theory expressed in terms of Majorana fermions and real scalar fields again allows additional terms which further break an Ising symmetry associated with the real scalar field. 
Near the phase transition, non-analytical cusps develop in the grand potential and thermodynamic quantities.
Due to the symmetry breaking terms, the details of dynamics in our theory will be different from the minimum model of Majorana fermions coupled to a real field via Yukawa interactions, which can lead to the $\mathcal{N}=1$ supersymmetry 
\cite{Wess74,Friedan84, Fei17} when the mass of the real field is further tuned to zero. Supersymmetry has been considered as an interesting emergent phenomenon in various condensed matter systems \cite{Fendly03,Thomas05,Lee07,Yue10,Grover14,Jian15,Balents98}. 

The topological phase transitions studied here belong to a class of critical points characterized by free Majorana fermion dynamics.
The unique bulk signatures of these freely propagating Majorana fermions are the focus of this study. We further illustrate how interactions between Majorana fermions and other fluctuating fields will affect the non-analytical cusps near phase transitions.
Interestingly, certain gradient operators (see section \ref{cusps}) which usually lead to drastic changes of global topology do not contribute to the cusp structures at all, not even in a quantitative way, although interactions always modify the non-analytical structures obtained in the fixed point theory. These non-analyticities hence can be an indicator of interactions between Majorana fermions and fluctuations.

Apart from the motivation of understanding topological phase transitions, let us also emphasize another motivation of studying thermodynamic signatures of Majorana fermions at topological phase transitions from a more pragmatic point of view.
Although Majorana fermionic modes are believed to be present in many condensed matter systems, 
experimental studies have been mainly focused on zero energy Majorana modes or bound states at interfaces.  
The thermodynamic signatures obtained below suggest a potentially complimentary approach where the Majorana mass can be tuned by varying parameters in our many-body systems such as interaction strength and/or fermion densities. The Majorana zero energy mode can further proliferate into the bulk at the critical point, a remarkable property of fine tuned strong interactions.

In the following, we are going to present the results of our investigation on the topological phase transition which further breaks the $U(1)$ gauge symmetry, rotation symmetry, etc.
In section II, we provide a pedagogical introduction to Majorana physics in superfluids as a cure to the redundancy of many-body descriptions of fermionic modes in superfluids.
In section III, we present the effective theory and derive the fixed point Hamiltonian for the topological phase transitions via the renormalization group equation approach
in the standard Nambu representation for superfluids. Emphasis will be made on the effect of spontaneous symmetry breaking on the phase transitions.
In section IV, we specify the nature of cusps near the phase transitions in 2D. The fixed point theory suggests that the transitions in $d$-dimensional isotropic superfluids shall be $(d+1)$th order. This conclusion is valid for both time reversal symmetry breaking and time-reversal symmetric superfluids. In addition, we show that the cusp structure is robust against the change of UV physics while the topology is not. 
In section V, we study the effects of interactions between Majorana fermions and fluctuations in superfludis. These interactions do not change the order of phase transitions but can modify the shape of the cusps. 
In section VI, we reformulate our theory in terms of Majorana fermions, instead of Nambu fermions, interacting with a real scalar field. We compare this with the conventional field theory model of this type. 
The action for the real scalar field explicitly breaks the usual $Z_2$ symmetry due to the breaking of $U(1)$ symmetry both in superfluid phases and at transitions. We also discuss the relation between these topological transitions and supersymmetry conformal field theories.
In section VII, we wrap up our investigation by discussing implications on physical observables and possible experimental detection of cusps at these transitions.
In section VIII, we conclude our research and identify a few open questions that we plan to pursue in the future.

\section{General phenomenology of Majorana fermions in superfluids}

In general, Majorana fermions can emerge naturally in superconductors or fermionic superfluids as a result of $U(1)$ symmetry breaking.
The creation of a quasi-particle with positive energy $E$ is always equivalent to the annihilation of a quasi-particle with negative energy $-E$.
This makes the fermionic superfluids or superconductors a potential condensed matter platform to study Majorana fields first pointed out long ago in relativistic field theory where particles are their own antiparticles.
The redundancy between creation and annihilation of fermionic states in the standard many-body representation of a superfluid/superconducting state has been well known.
Only relatively recently its intimate relation to Majorana fields and Majorana fermions started to be scrutinized quite intensively, partly because of the accessibility to fermions with relativistic-like characters. 
During the last decade or so, there have been very intensive experimental activities aiming at creating Majorana zero energy modes on the surface of various quantum materials and 
in hybrid structures, and using condensed matter systems as a powerful platform to investigate Majorana particles that have never been unambiguously verified to exist in high energy particle physics \cite{Alicea12,Beenakker13,Elliot15}.

There are a few main trains of thoughts that have led to the extensive studies of Majorana fermions in condensed matter physics.
One is related to early theoretical exploration of fractional quantum Hall states which reveals a close relation between the non-Abelian statistics and Majorana modes residing in vortices \cite{Moore91,Read96,Nayak96}. More recently 
strongly interacting Majorana fermions in vortex lattices were also discussed in Ref. \cite{Rahmani15,Affleck17}. 
The other one is from the consideration of quantum computation where neutral Majorana particles not only enjoy an advantage of immunity from the conventional decoherence effects, but their braiding also leads to potential universal quantum gates needed for quantum computing \cite{Kitaev03,Nayak08,Freedman02}.
The one that most directly triggered many recent activities is perhaps the idea of creating Majorana modes via a topological insulator that had become available in many laboratories \cite{Fu08}. This idea was later generalized to much broader classes of interfaces, surfaces and hybrid structures \cite{Alicea12,Beenakker13}.  

The emergence of Majorana physics in superfluids or superconductors can be clearly seen by recalling the construction of quasi-particle space.
In non-interacting systems, at each given wave vector ${\bf k}$, there is only one pair of creation and annihilation operators $(c_{\bf k}^\dagger, c_{\bf k})$ where $\{ c^\dagger_{\bf k},
c_{\bf k}\}=1$.
If one is interested in the excitations out of the Fermi sea, it is more convenient to separate the degrees of freedom below and above the Fermi surface.
Then one can introduce $(d^\dagger_{\bf k}, d_{\bf k})$ for $|{\bf k}| < k_F$ so that $d_{\bf k}=c_{-\bf k}^\dagger$, $d_{\bf k}^\dagger=c_{-\bf k}$, and $\{d^\dagger_{\bf k},d_{\bf k}\}=1$;
$d^\dagger_k$ is the creation operator of a hole-like state.
And for $|{\bf k}| > k_F$, one keeps the notion of $(c_{\bf k}^\dagger, c_{\bf k})$. (We mute the spin indices here as they do not carry significance for our purpose.)

\begin{figure}
\includegraphics[width=\columnwidth]{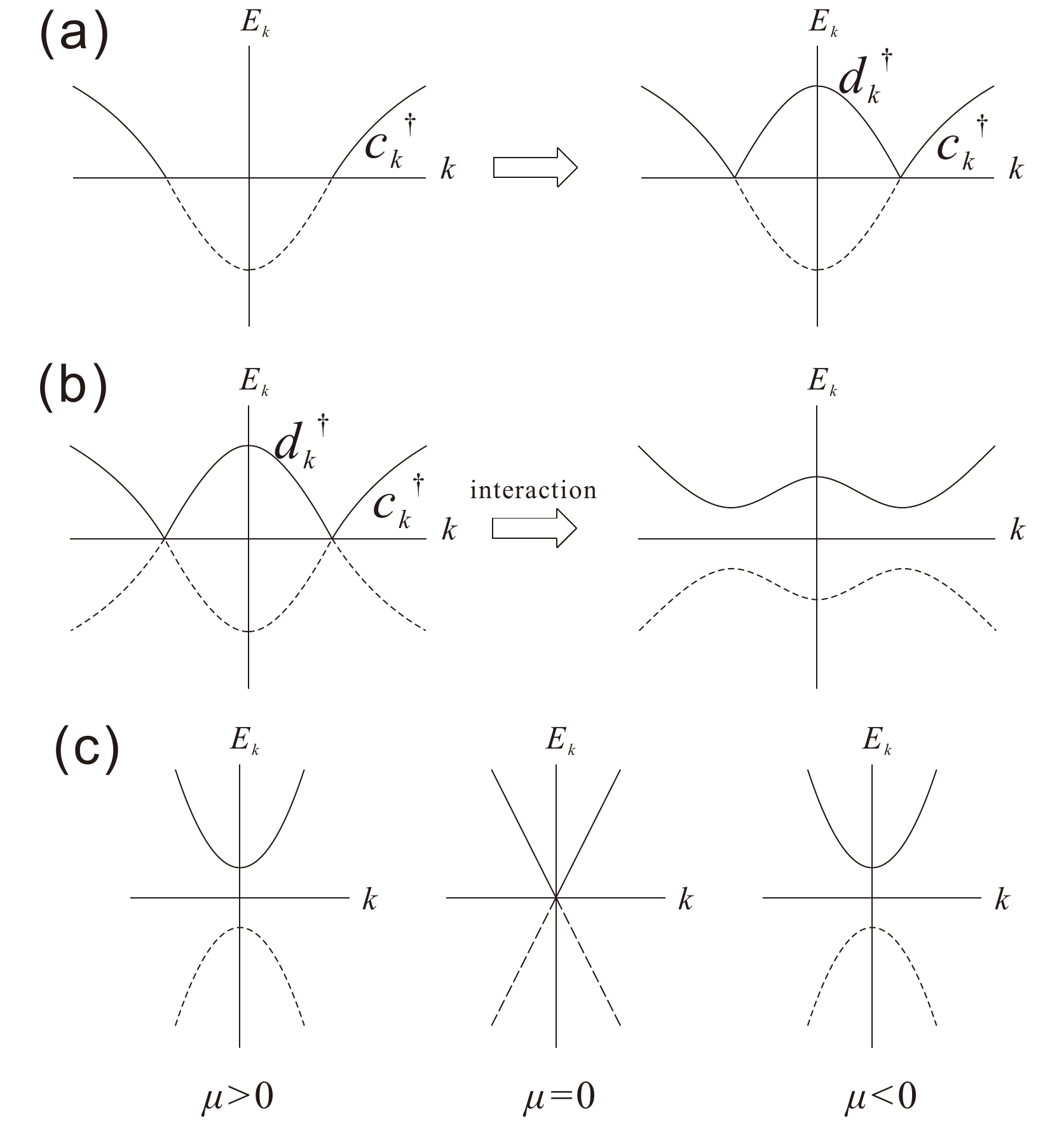}
\caption{(a) Two equivalent ways to describe non-interacting fermions. $E_k$ is the energy measured from chemical potential $\mu$. 
In the left figure, there is one complex fermion degree of freedom at each ${\bf k}$ associated with the creation operator $c^\dagger_{\bf k}$. 
All states below the chemical potential with negative energy are filled (dashed line), while all states above chemical potential with positive energy are empty (solid line). 
In the right figure, for the states below the chemical potential (dashed line) one redefines the fermion creation operator as an annihilation operator for a hole-like fermion so that $c^\dagger_{\bf k}=d_{-\bf k}$.  $d^\dagger_{\bf k}$ defined in this way and $c^\dagger_{\bf k}$ for the rest of momentum space only create excitations with positive energies.
(b) One can extend the definition of $c^\dagger_{\bf k}$ and $d^\dagger_{\bf k}$  to the whole momentum space to preserve the particle-hole symmetry at each given ${\bf k}$ and include full positive and negative energy bands. Note that now at any given ${\bf k}$, there appear to be two complex fermion degrees of freedom, $d^\dagger_{\bf k}$ and $c^\dagger_{\bf k}$. Hence, the representation is redundant. 
When interactions are introduced, these two bands can be hybridized as shown on the right. The redundancy can be removed by either removing the lower band so that for each given ${\bf k}$ there is only one single complex fermion degree of freedom, or by keeping only half fermion or one Majorana degree of freedom in lower and upper bands  respectively.
(c)The phase transition driven by chemical potential. For $\mu<0$ the system is in the gapped topologically trivial superfluid phase. At $\mu=0$ the gap closes. For $\mu>0$, the bulk gap reopens (edge states are not shown here) and the system is in the topological superfluid phase. }\label{Nambu}
\end{figure}

As a superfluid or superconductor breaks $U(1)$ symmetry which leads to the hybridization of $c^\dagger_{\bf k}$ and $c_{-\bf k}$ (or $d^\dagger_{\bf k}$) for arbitrary ${\bf k}$, one needs to extend the definition of $(c^\dagger_{\bf k}, c_{\bf k})$ and $(d^\dagger_{\bf k}, d_{\bf k})$ to all  possible ${\bf k}$. 
This results in an effective Nambu space where at each ${\bf k}$ there exist two pairs of creation and annihilation operators $(c^\dagger_{\bf k}, c_{\bf k})$ and $(d^\dagger_{\bf k}, d_{\bf k})$ instead of the one pair $(c^\dagger_{\bf k}, c_{\bf k})$ we start with.
This is of course redundant.
The redundancy is then removed by the constraint that the creation of a particle-like excitation of momentum ${\bf k}$ in superfluids or superconductors has to be equivalent to the creation of a hole-like state and vice versa.  
Indeed, together with relativistic characteristics of fermions due to either underlying band twisting \cite{Fu07,Moore07,Fu08} or odd-parity pairing, it can lead to Majorana fermions.  See Fig. \ref{Nambu} for illustration.  
In fact, this is what happens in $p$-wave superfluids considered below, where the symmetry between particle-like and hole-like states, together with the emergent relativistic fermion dispersion, results in Majorana fermions.

In the following discussions, we will then utilize the Nambu fermion representation to carry out the discussion as this naturally arises in the course of many-body state constructions.
The Nambu fermion representation is also the most convenient if one is interested in making a connection between the effective theory and the original UV model and
perhaps microscopically what happens to the superfluids at topological phase transitions.
Later, this is shown to be equivalent to the irreducible Majorana representation where the Majorana fields can be reconstructed out of the Nambu fermions. The effective interactions between Majorana fermions and the real scalar field can be mapped back to the theory in terms of Nambu fermions.

\section{Effective theory for topological phase transitions with broken symmetries}

The main difference between the topological phase transitions we are interested in here and many other topological phase transitions discussed before is that the transition occurs at a point that exhibits spontaneous symmetry breaking. 
This is because the two phases on the two sides of the transition have the same conventional local ordering and same local pairing order parameters
but only differ in topology. The effective theory in the Nambu fermion representation should fully represent this fact and allow additional symmetry breaking terms explicitly to accommodate this unique feature.
For simplicity but without loss of generality, we first focus on isotropic superfluids (i.e. superfluids with isotropic gaps) of 2D spinless fermions.
Later, we extend the analysis to other isotropic superfluids with time reversal symmetry (TRS); the cusp structures are robust, with or without TRS.

To start with, we introduce spinless Nambu fermions in 2D as
\begin{equation}
\psi({\bf k}) =(c_{\bf k}, d_{\bf k})^T,\; \psi^\dagger({\bf k})=\tau_x\psi(-{\bf k}),
\end{equation}
where the second equation follows the definition of $d_k$: $d_{\bf k}^\dagger=c_{-{\bf k}}$ and $c^\dagger_{\bf k}=d_{-\bf k}$, and $\tau_x$ is the Pauli matrix in Nambu particle-hole space.
This, together with a single pairing amplitude dynamic field (or real scalar field) resulted from the spontaneous breaking of  $U(1)$ (or $SO(2)$ rotational) and time reversal symmetries, forms the basis of the effective theory near the transition. 
Below we discuss the renormalization flows of the effective field theory and their consequences on the universality.
The renormalization group (RG) analysis suggests the fixed point Hamiltonian in the infrared limit relevant to the observables.

\subsection{Effective field theory}

Near the transition in the low energy limit, both the real bosonic and the fermionic fields shall develop an emergent relativistic dispersion. 
We introduce the effective Lagrangian density in the Euclidean space for 2D spinless $p_x-ip_y$ type superfluid  as
\begin{eqnarray}\label{Lagrangian}
&&\mathcal L= \mathcal L_{\phi} +\mathcal L_{\psi} +\mathcal L_{\psi\phi} +\mathcal L^{SSB}_{\{\psi,\phi\}};
\nonumber \\
&& 
\mathcal L_{\phi}=-\frac12\phi (\partial^2_\tau+v_b^2\nabla^2-m^2)\phi+g_b \phi^4; \nonumber\\
&&\mathcal L_{\psi}=\frac12\overline\psi[\partial_\tau+v_f{\vec\tau}\cdot {(-i{\bf\nabla})}-\mu\tau_z]\psi; \nonumber \\
&&\mathcal L_{\psi\phi} = g_{bf}\phi^2\overline{\psi}{\bf \tau}_z\psi; \nonumber\\
&& \begin{split}\mathcal L^{SSB}_{\{\psi,\phi\}} =& \lambda_{bf}\phi\overline\psi\tau_z\psi+\lambda_b\phi^3.
\end{split}
\end{eqnarray}
Here $\overline\psi$ ($\psi$) is the Grassmann field for Nambu fermions $\psi^\dagger$ ($\psi$) defined before. $\phi$ is a real scalar or a real bosonic field coupled to the neutral Nambu or Majorana field.
$\vec\tau=(\tau_x,\tau_y)$, where $\tau_{x,y,z}$ are Pauli matrices in Nambu particle-hole space, and $\nabla$ is the 2D gradient operator.
$\mathcal L_{\phi} (\mathcal L_{\psi})$ describes the fluctuating bosonic (fermionic) field dynamics, while $\mathcal L_{\psi\phi}$ describes the most relevant gauge invariant interactions
between the two fields. $\mathcal L_{\phi}$, $\mathcal L_{\psi}$, $\mathcal L_{\psi\phi}$ are all invariant under a $\pi$ phase shift of the $\phi$ field, i.e. under $\phi \rightarrow - \phi$.
$\mathcal L^{SSB}_{\{\psi,\phi\}}$ describes additional interactions that can also be attributed to spontaneous symmetry breaking (SSB). Terms in $\mathcal L^{SSB}_{\{\psi,\phi\}}$ are usually absent if the phase transition occurs without breaking the underlying $U(1)$ gauge symmetry; however they shall naturally appear when the symmetry is broken as in superfluid phases we are considering (see below for more discussions.)

One can easily read out the bare Green's functions from the effective Lagrangian 
\begin{equation}\label{boson}
G_B^{(0)}(\omega,k)=-\frac1{\omega^2+v_b^2k^2+m^2},
\end{equation}
\begin{equation}\label{fermion}
{G}_F^{(0)}(\omega,k)=-\frac{i\omega+v_f{\vec\tau}\cdot{\bf k}-\mu\tau_z}{\omega^2+v_f^2k^2+\mu^2}.
\end{equation}

We make the following remarks on the effective theory before proceeding further.

1) In the effective theory, the real scalar field $\phi$ describes pairing amplitude fluctuations in a symmetry breaking state and the Nambu field $\psi$ represents emergent Majorana fermions near transitions.  We have introduced the effective chemical potential $\mu$ which plays the role of Majorana mass \cite{mu}.
We have also reduced the microscopic complex pairing field to a real scalar field as the Majorana fermions are real and the coupling to phase fluctuations can be shown to be infrared irrelevant.
 The Goldstone mode hence has been muted in the effective model and we restrict ourselves to the interactions with the amplitude modes.
The condensed part of the pairing field has been treated as a classical field and thus is not explicitly present in the effective model. In fact, since the condensed part is not present in the fluctuating field $\phi$, we should have  $\langle\phi \rangle=0$.

2) The pair condensate that leads to SSB, though not shown explicitly in the effective theory, manifests in two ways. One is to suggest relativistic features of fermionic fields $\psi$, consistent with a $p_x-ip_y$ pair condensate. 
Physically, fermion velocity $v_f$ is related to the microscopic condensate amplitude $\Delta$. $\phi$ fields are also in an explicit relativistic form.
The relativistic feature is fully consistent with the pair fluctuations induced by the relativistic fermions as well as the particle-hole symmetry of pairs. 
The other effect of the condensate is to introduce symmetry breaking terms in the interaction action which we elaborate on below.

3) In the effective theory, we only include the most relevant interactions constructed out of the two fundamental fields $\psi$ and $\phi$.
In $\mathcal L^{SSB}_{\{\psi,\phi\}}$, we further include terms that naturally emerge in symmetry breaking superfluids.
Originating from the breaking of the $U(1)$ gauge symmetry at the critical point, these terms also break a discrete symmetry under a transformation of $\phi \rightarrow -\phi$.
Here, we do not show interactions involving gradient operators and  terms with more than five field operators; they are less relevant and do not affect the discussions 
we have later on. 

4) As a field theory model, in the absence of $\mathcal L^{SSB}_{\{\psi,\phi \}}$, Eq. (\ref{Lagrangian}) exhibits a few obvious fixed points where scale invariance emerges in the infrared limit. These fixed points can be identified as critical points for different physical systems. The first one is the standard Wilson-Fisher fixed point of the real bosonic field when the mass $m$ is fine tuned.  
Another is the free Nambu fermion fixed point, or more precisely Majorana fermion fixed point as illustrated more explicitly later in the Majorana representation, when the fermion chemical potential is fine tuned.
 The third less obvious one can be related to a multi-critical supersymmetry fixed point when the Yukawa interaction in $\mathcal L^{SSB}_{\{\psi,\phi\}}$ is included. 
 However, it requires high degrees of fine tuning in our case, much more than in generic critical phenomena. We do not proceed along this direction in this article, but will nevertheless briefly comment on it later in the Majorana fermion representation where the issue of supersymmetry becomes more apparent. 

5) In this low-energy effective theory, the ultra-violet (UV) cut-off $\Lambda_c$ is much smaller than the UV cut-off of the microscopic model.
For a generic situation, $m^2$ and the interaction constants $g_b$ and $\lambda_b$ of the fluctuating fields, are set to be positive. 
The cusp structure in thermodynamic quantities is purely dictated by infrared physics, and naturally its leading order behaviour does not rely on our detailed assumption of $\Lambda_c$.
To facilitate our discussions on interactions and interacting Majorana fermions, we have also set $\Lambda_c \ll \Delta$ and hence  much less than the mass $m$  as the magnitude of $m$ shall be related to microscopic condensate amplitude $\Delta$.  
This is for the convenience of the discussion, but the qualitative effects of the interactions do not rely on this particular regularization scheme.
A microscopic model might result in negative $g_b$ implying a discontinuous first order phase transition. We do not consider such a possibility here and restrict ourselves to continuous transitions where $g_b$ is positive.

\subsection{A renormalization-group-equation (RGE) analysis via Nambu fermions}

As the topological phase transition occurs inside a symmetry breaking state, generically without further fine tuning, the bosonic fluctuation field is always massive and drives the renormalization group (RG) flows away from the fine tuned Wilson-Fisher fixed point
in the infrared (IR) limit. 
For this reason, instead of rescaling all interactions with respect to the massless free field theory,
it is most convenient to rescale all the operators in the effective field theory with respect to the mass term $m$ of the real scalar field. 
As the mass term of the boson field dominates and is most relevant in the infrared limit, we demand $m^2$ not rescale during renormalization when the UV cut-off $\Lambda_c \ll m$, i.e.
setting the following scaling dimensions for the two fields:
\begin{align}\label{scaling}
d[\psi]&=d/2,&& d[\phi]=(d+1)/2,
\end{align}
where $d[.. ]$ denotes the scaling dimension of $[..]$, and $d$ is the spatial dimension of the system. 
The scaling dimension of $\phi$ is different from that of the standard RG where the UV cut-off $\Lambda_c$ is much larger than the mass $m$ in the theory and the massless part of  the theory remains invariant under rescaling with $d[\phi]=(d-1)/2$. As a result, the scaling dimensions of each operator are also different from those in the standard RG: 
\begin{eqnarray}
&& d[m]=0,\; d[\mu]=1,\;  d[\lambda_{bf}]=-(d-1)/2,  \nonumber \\
&& d[g_{bf}]=-d,\; d[\lambda_b]=-(d+1)/2,\;d[g_b]=-d-1. 
\end{eqnarray}
Note that in 1D the Yukawa interaction $\lambda_{bf}$ is marginal. The physical dimensions of each operator in the relativistic effective model are $[m]=L^{-1}$, $[\mu]=L^{-1}$, $[g_{bf}]=L^{d-2}$, $[g_b]=L^{d-3}$, $[\lambda_{bf}]=L^{(d-3)/2}$ and $[\lambda_b]=L^{(d-5)/2}$. 
We summarize these results and compare the difference between standard RG and the RG used in the SSB state in table \ref{dimensions}.

\begin{table}\label{dimensions}
\center
\begin{tabular}{|p{1.8cm}|p{1.7cm}|p{2.2cm}|p{2.2cm}|}
\hline
parameters & physical dimension for relativistic dispersions& standard scaling dimension i.e. for $m\ll\Lambda$ & SSB phase scaling dimension for $m\gg\Lambda$\\
\hline
$\phi$ & $L^{-(d-1)/2}$ & $(d-1)/2$ & $(d+1)/2$\\
\hline
$\psi$ & $L^{-d/2}$ & $d/2$ & $d/2$\\
\hline
$v_b$ & $L^0$ & 0&$-1$\\
\hline
$v_f$ & $L^0$ & 0 & 0\\
\hline
$m$ & $L^{-1}$ & 1 & 0\\
\hline
$\mu$ & $L^{-1}$ & 1 & 1\\
\hline
$g_b$ & $L^{d-3}$ & $3-d$ & $-d-1$\\
\hline
$g_{bf}$ & $L^{d-2}$ & $2-d$ & $-d$\\
\hline
$\lambda_b$ & $L^{(d-5)/2}$ & $(5-d)/2$ & $-(d+1)/2$\\
\hline
$\lambda_{bf}$ & $L^{(d-3)/2}$ & $(3-d)/2$ & $-(d-1)/2$\\
\hline
\end{tabular}
\caption{Comparison of the scaling dimensions of each operator in two renormalization schemes. The physical dimensions of each operator are obtained from the standard relativistic Lagrangian where $[\omega]=[k]=L^{-1}$ and $[v_b]=[v_f]=L^0$. In the standard renormalization scheme where $m\ll\Lambda_c$, the scaling dimension is negative of the physical dimension. In contrast, in the  renormalization scheme used for spontaneous symmetry breaking (SSB) states discussed here where $m\gg\Lambda_c$, $m$ is subject to  almost no renormalization effects (see text for more discussions). This suggests a more economical alternative scale transformation that leaves the mass term invariant and redefines the scaling dimensions of other operators accordingly.}\label{dimensions}
\end{table}

As we intend to apply the effective field theory to the infrared energy scale much lower than $m$, we can expand the boson propagator in orders of the small parameter $1/m^2$ and approximate the boson propagator to the leading order of this parameter
\begin{equation}\label{boson}
G_B^{(0)}(\omega,k)\approx-\frac1{m^2}\left(1-\frac{\omega^2+v_b^2k^2}{m^2}\right)\approx-\frac1{m^2}.
\end{equation}
In this approximation, the boson propagator does not explicitly depend on frequency or momentum. 
As a result, in the leading order of $1/m^2$ there are no renormalization effects on field operators $\phi$ and $\psi$, or velocities $v_b$ and $v_f$.
The self-energies only renormalize the effective mass $m$ and effective chemical potential $\mu$.

To analyze the renormalization flows, it is convenient to define the following dimensionless operators: 
\begin{eqnarray}
&& \tilde\mu=\mu\Lambda^{-1}, \nonumber\\
&&  \tilde \lambda_{bf}=\lambda_{bf}\Lambda^{(d-1)/2}/m,\;\tilde g_{bf}= g_{bf}\Lambda^d/m^2 \nonumber\\
&&\tilde\lambda_b=\lambda_b\Lambda^{(d+1)/2}/m^3,\;   \tilde g_b=g_b\Lambda^{d+1}/m^4.%\; 
\end{eqnarray}
Here we incorporate $m$ into the definition of the dimensionless parameters to compensate the difference between physical and scaling dimensions of each operator. 
This definition has two benefits: (1) the RG equations of the interaction constants do not contain $m$ explicitly; (2) all the above dimensionless interaction constants can still be treated as small parameters even if the physical interaction strengths normalized in the standard way in terms of the UV cutoff reach the order of unity. 

\begin{figure}
\includegraphics[width=\columnwidth]{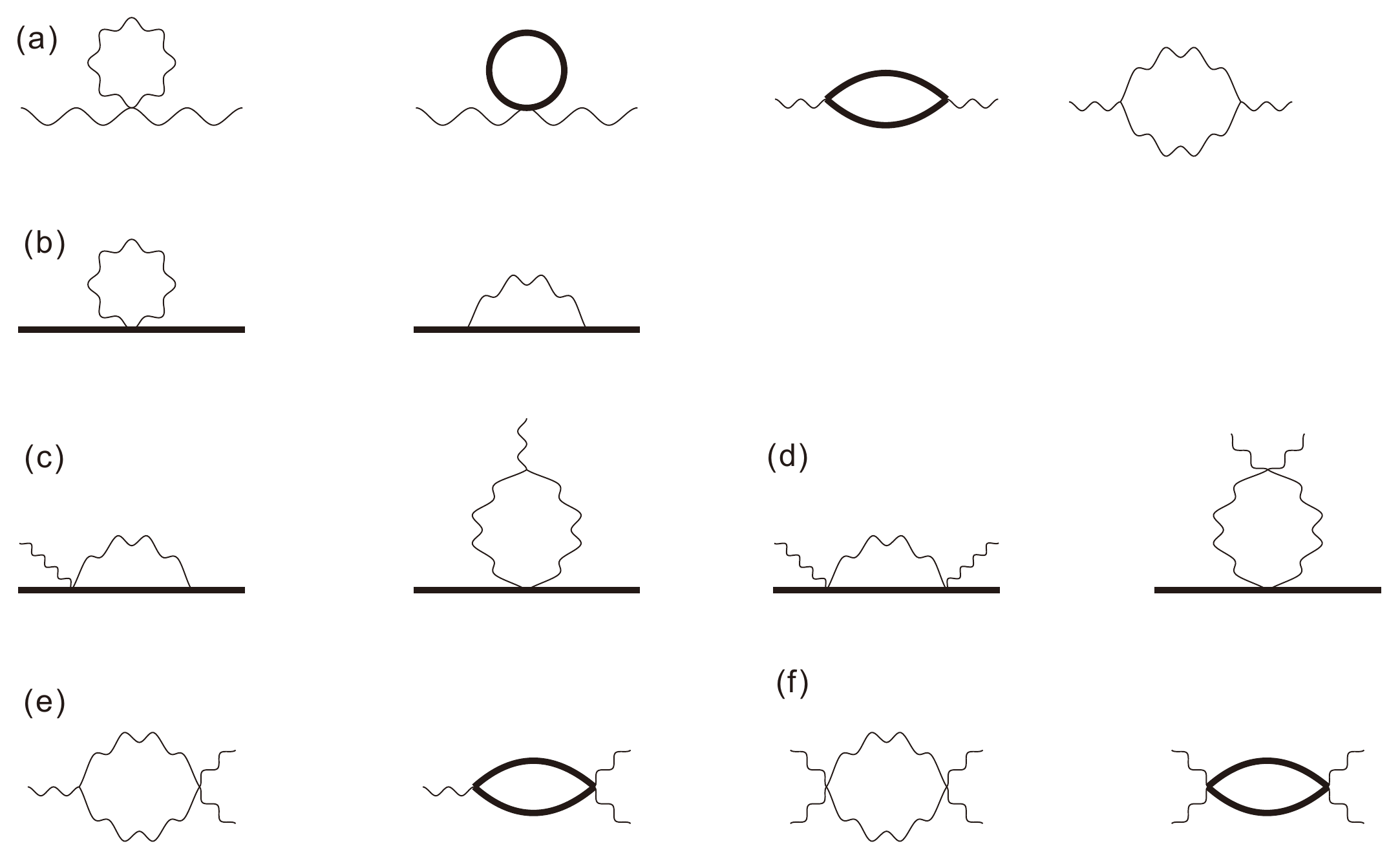}
\caption{\label{RGdiagram} Diagrams included in our perturbative RG analysis. We keep the leading order contributions up to second order of interaction constants and up to first order of $\mu$. The bold solid lines represent fermion propagators and the wavy lines boson propagators. (a) boson mass renormalization, (b) effective fermion chemical potential renormalization (c) $\lambda_{bf}$ renormalization (d) $g_{bf}$ renormalization (e) $\lambda_b$ renormalization (f) $g_b$ renormalization.}
\end{figure}

In the following discussions of Nambu fermion representation, we set $d=2$.
We also keep leading order contributions up to the second order of interactions in the RG equations;
in addition, since we are interested in the phase transition where $\tilde{\mu}$ is small, we include its effect by keeping diagrams linear in $\mu$. In Fig. \ref{RGdiagram}, we show the complete set of diagrams following these criteria.

\begin{figure}
\includegraphics[width=\columnwidth]{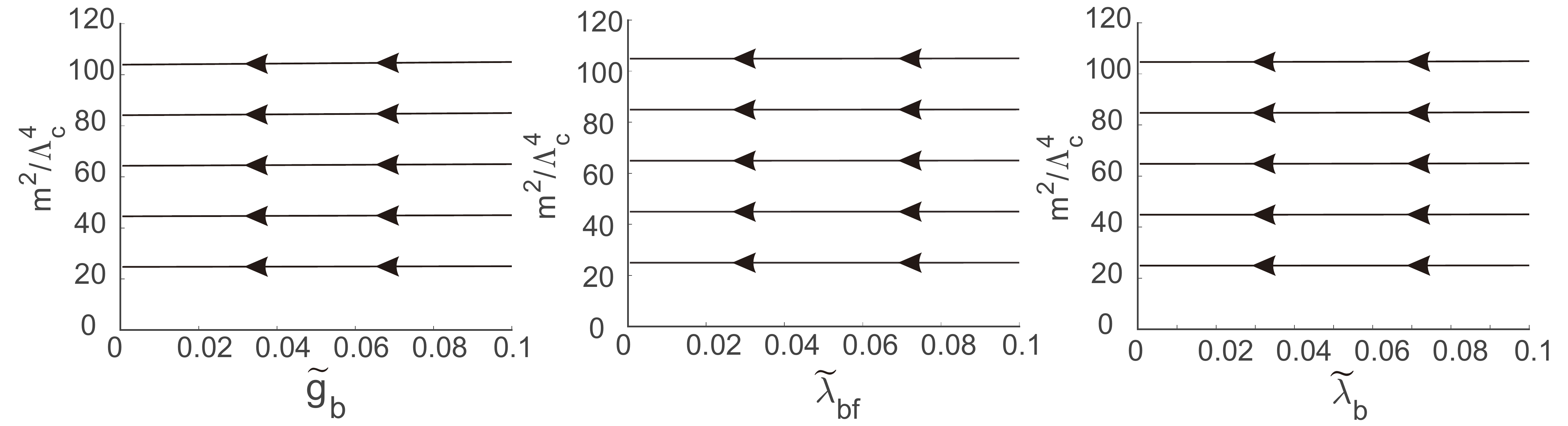}
\caption{RG flows in the reduced $m^2-\tilde g_b$, $m^2-\tilde\lambda_{bf}$  and $m^2-\tilde\lambda_{b}$ plane. Arrows indicate the direction of flow towards the infrared direction. When the UV scale $\Lambda_{c}$ of the effective theory is much less than $m$, 
$m(\Lambda)$ itself receives almost no renormalization effects and remains approximately invariant under the change of running momentum scale $\Lambda<\Lambda_c$ [see Eq. (\ref{mass})] suggesting an effective {\it disordered fixed point}.}\label{massRG}
\end{figure}

To verify that the system is indeed in the $U(1)$ symmetry breaking phase and justify Eq. (\ref{boson}), we first analyze the renormalization of mass $m$ under rescaling $\Lambda\to\Lambda e^b$, $b<0$ using the RG equation
\begin{equation}\label{mass}
\frac{dm^2}{db}=K_{d+1}^{-1}v_f^{-1}(-2\tilde\mu\tilde g_{bf}-6v_f^2\tilde g_b+2\tilde\lambda_{bf}^2+18v_f^2\tilde\lambda_{b}^2)m^2,
\end{equation}
where $K_{d+1}=2^d\pi^{(d+1)/2}\Gamma\left(\frac{d+1}2\right)$, and $\Gamma$ is the Gamma-function. 
This equation describes the slow flow of $m^2$ in the limit $\Lambda\to0$ (Fig. \ref{massRG}). In the IR regime, $m^2$ is approximately a constant under rescaling and the system remains in the symmetry breaking phase.
$m^2$ can have arbitrary values depending on the initial condition.

Next, we obtain the RG equations of the effective chemical potential and interaction constants to study the topological phase transition: 
\begin{equation}
\frac{d\tilde \mu}{db}=-\tilde\mu+K_{d+1}^{-1}v_f^{-1}(v_f^2\tilde g_{bf}+\tilde\mu\tilde\lambda_{bf}^2),
\end{equation}
\begin{equation}
\begin{split}
\frac{d\tilde\lambda_{bf}}{db}=&\frac{d-1}2\tilde\lambda_{bf}-3K_{d+1}^{-1}v_f^{-1}\tilde\lambda_{bf}\tilde g_{bf}\tilde\mu\\
&+K_{d+1}^{-1}v_f(3\tilde\lambda_{bf}\tilde g_b+12\tilde\lambda_b\tilde g_{bf}),
\end{split}
\end{equation}
\begin{equation}
\frac{d\tilde g_{bf}}{db}=d\tilde g_{bf}+30K_{d+1}^{-1}v_f\tilde g_{bf}\tilde g_b-2K_{d+1}^{-1}v_f^{-1}\tilde g_{bf}^2\tilde\mu,
\end{equation}
\begin{equation}
\begin{split}
\frac{d\tilde \lambda_b}{db}=&\frac{(d+1)}2\tilde \lambda_b+3K_{d+1}^{-1}v_f^{-1}\tilde g_{bf}\tilde\lambda_b\tilde\mu\\
&+K_{d+1}^{-1}v_f^{-1} (81v_f^2\tilde g_b\tilde\lambda_b+4\tilde\lambda_{bf}\tilde g_{bf}),
\end{split}
\end{equation}
\begin{equation}
\begin{split}
\frac{d\tilde g_b}{db}=&(d+1)\tilde g_b+4K_{d+1}^{-1}v_f\tilde g_{bf}\tilde g_b\tilde\mu\\
&+K_{d+1}^{-1}v_f^{-1}(84v_f^2\tilde g_b^2+2\tilde g_{bf}^2).
\end{split}
\end{equation}

\begin{figure}
\includegraphics[width=\columnwidth]{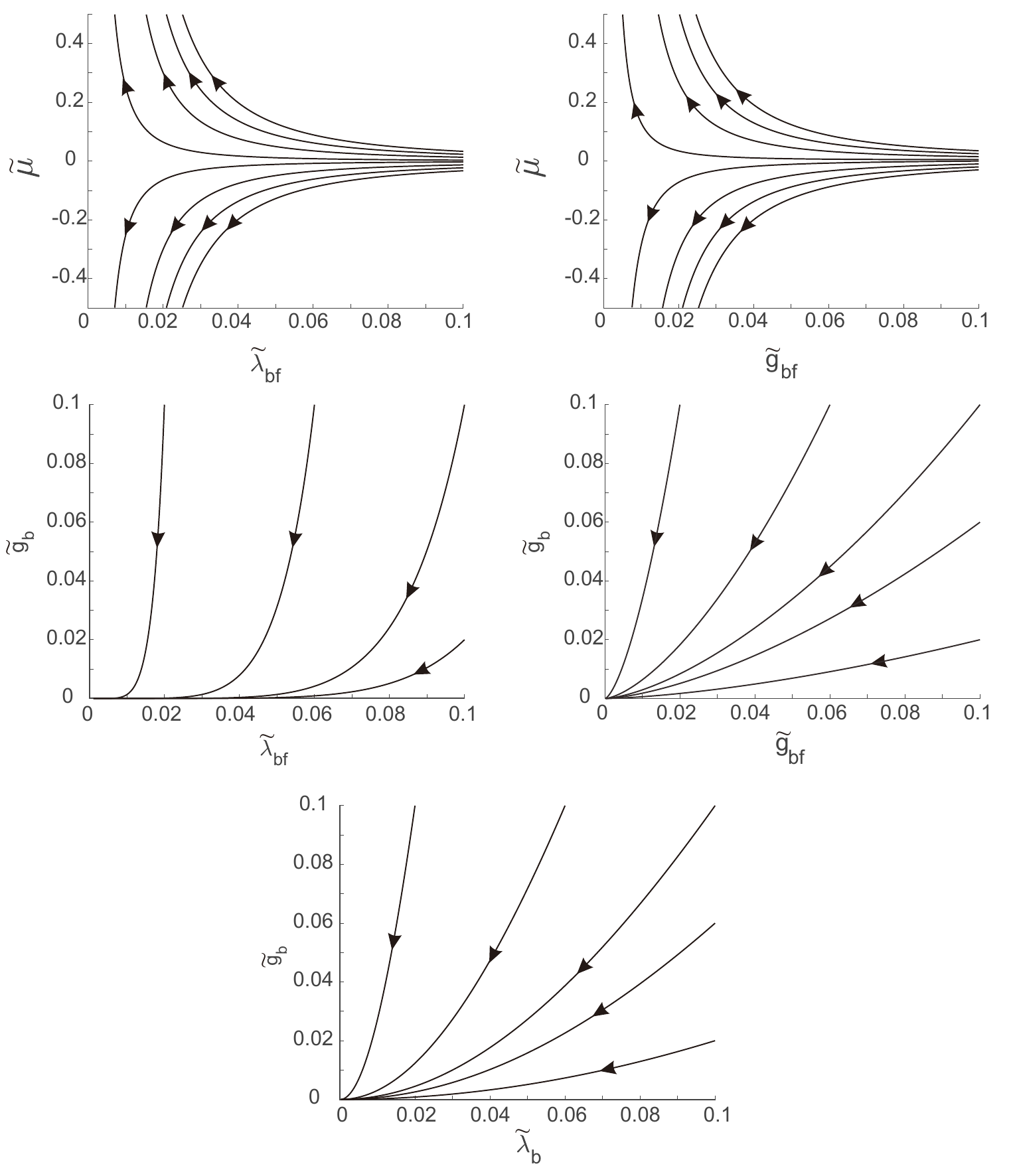}
\caption{RG flows of effective chemical potential $\mu$ and interactions for 2D chiral superfluids.
We select two operators for each figure and plot the flows in their reduced plane by setting other interactions to zero. All the interactions flow to zero in the infrared limit and $\mu$ has an unstable IR fixed point at $\mu=0$. This illustrates that all interactions are irrelevant; while the effective chemical potential is a relevant operator that drives the topological phase transition.} \label{RGflow}
\end{figure}

We analyze the fixed point of these coupled differential equations by numerically plotting the flows of these operators under rescaling (Fig. \ref{RGflow}). As one can see, all the interactions flow to zero in the IR limit suggesting the system is effectively non-interacting at the critical point. The effective chemical potential has an unstable fixed point $\tilde\mu=0$, indicating $\mu$ is the only relevant operator that drives the phase transition. 

In conclusion, the IR stable fixed point corresponds to $\tilde\lambda_{b}=\tilde\lambda_{bf}=\tilde g_b=\tilde g_{bf}=0$ but with arbitrary $m$; the fluctuating boson field is effectively at a disordered fixed point and the system is weakly interacting near the critical point. 
The effective fermion chemical potential $\mu$ is the only relevant operator with IR unstable fixed point $\tilde\mu=0$, which drives the phase transition between topological and non-topological superfluids.
Below we will focus on the properties near such a fixed point.

\subsection{Identification of the fixed point action for topological phase transitions}\label{fixedpointH}

The RG analysis above indicates that the topological phase transition corresponds to free Nambu  or Majorana fermions weakly interacting with massive bosons, with the weak interaction limit as a stable fixed point in the hyper-surface of  $\tilde\mu=0$, and a fixed point Lagrangian
\begin{equation}\label{fpl}
\mathcal{L}_{\text{FP}}=\frac12\overline\psi[\partial_\tau+v_f{\vec\tau}\cdot (-i\nabla)-\mu\tau_z]\psi.
\end{equation}
We can then take this as a starting point to explore the physical consequences of weakly interacting Majorana fermions. 
The relevance of free Majorana fermions to the phase transition has been previously suggested in the course of studying duality theories. 
We refer to Ref. \cite{Senthil04a,Senthil04b,Seiberg16,Metilitski17,Wang17} for detailed discussions on dual representations.

As all interactions are renormalized to a weak coupling fixed point in the infrared limit and scale as $1/m^\alpha$, $\alpha=1,2,3,4...$,  
to understand the infrared physics we expand the effective action near the fixed point via a $1/m$ expansion. By keeping terms of up to $O(1/m)$ only, we obtain
\begin{equation}\label{FPaction}
\begin{split}
\mathcal L_{\text{IR}}=&
 \frac12\overline\psi[\partial_\tau+v_f{\vec\tau}\cdot (-i\nabla)-\mu\tau_z]\psi\\
&-\frac12\phi (\partial^2_\tau+v_b^2\nabla^2-m^2)\phi\\
&+\lambda_{bf}\phi\overline\psi\tau_z\psi,
\end{split}
\end{equation}
which describes a massive fluctuating  boson field and a Nambu fermion field interacting with each other.
We keep the boson field kinetic term to restore the relativistic feature but it does not play much a role in the following analysis of energetics.
Note as the Yukawa term above has the least negative scaling dimension among all the interaction operators,
the $1/m$ expansion here effectively selects out the most important contribution among all the infrared irrelevant interactions near the fixed point.
Below we explore the signatures near the topological phase transitions using this reduced form of the effective action.

\section{Cusps as signatures of free Majorana fermions emerging at a topological phase transition}\label{cusps}

In this section, we analyze the signatures of the topological phase transition in $p_x\pm ip_y$ superfluids.
We consider the grand potential $\Omega$ of the theory, especially near $\mu=0$ where the Nambu or Majorana fermions are massless, and expand $\Omega$ as a function of $\mu$.
We study the grand potential by a standard diagrammatic expansion and illustrate the origin of the non-analytical structures or cusps in the grand potential and compressibility. 

\subsection{Cusps}
Since the leading order non-analyticity at critical point is dictated by the IR physics, we can use the fixed point Lagrangian Eq. (\ref{fpl}) obtained from the RG analysis to find the leading non-analytical structure or cusp in the grand potential near the phase transition. The irrelevant operators can potentially modify the details of the cusps but do not affect the qualitative structure. We will discuss these irrelevant operators in the later sections. 

In principle, the fixed point Lagrangian can be solved exactly without diagrammatic expansions. 
However, the diagrammatic approach shows explicitly the origin of non-analytical terms and can later be generalized to incorporate the effect of interaction constants and other irrelevant operators.
  
\begin{figure}
\includegraphics[width=\columnwidth]{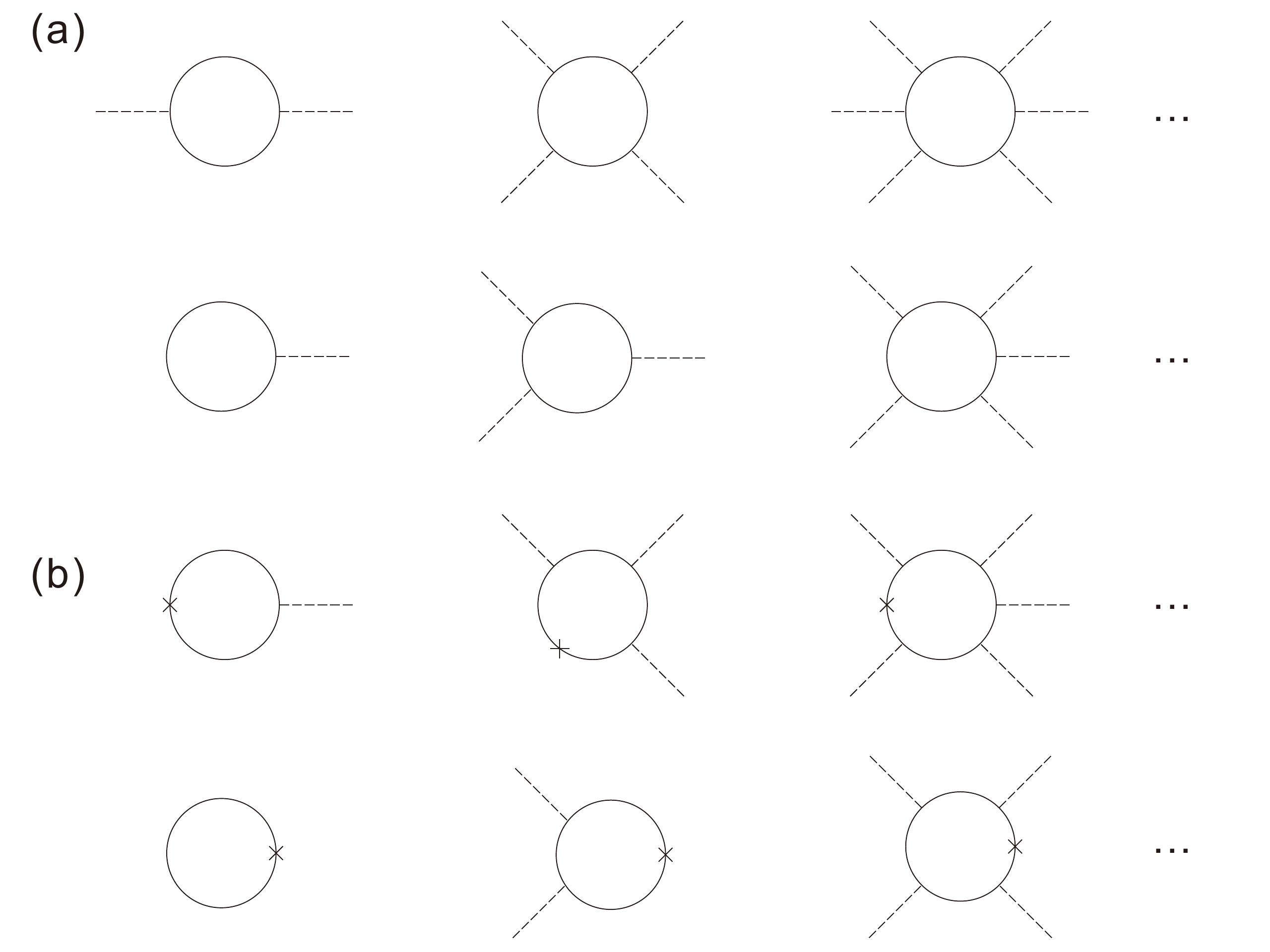}
\caption{Diagrams contributing to the grand potential $\Omega$.
(a) Diagrams for the fixed point action [see Eq. (\ref{fpl})], where all interactions are set to zero. The solid lines represent gapless free relativistic Majorana fermion propagators, while the dashed lines represent effective chemical potential $\mu$ which is treated as an external field. 
(b) When the gradient term associated with non-relativistic effects $\mathcal{L}_G$ [Eq. (\ref{gradient})] is included, diagrams shown here also contribute to the grand potential. The crosses represent the gradient term in $\mathcal{L}_G$, which is treated as a separate type of vertices.
In both (a) and (b), only diagrams with even number of fermion propagators contribute to the grand potential; those with odd number of fermion propagators vanish.
}\label{diagram}
\end{figure}

To find the non-analytical structures of the grand potential, it is convenient to treat $\mu$ as an external classical field and set $\mu=0$ in the bare fermion Green's function [Eq. (\ref{fermion})]. In this way, we can explicitly separate the non-analytical part from the analytical part systematically in arbitrary dimensions. 
Contributions to the grand potential involve all one-loop diagrams with external $\mu$-legs, such as those in Fig. \ref{diagram}(a). 

Under the fixed point Lagrangian $\mathcal L_{\text{FP}}$, the number of fermionic propagators is equal to the number of external legs in one-loop diagrams.
We notice that only diagrams with even number of fermion propagators, i.e. those are even orders of $\mu$, contribute to the grand potential; while diagrams with odd numbers of fermion propagators vanish identically. 
This is because each vertex is associated with a traceless matrix $\tau_z$. For the diagrams with odd number of fermion propagators, the product of all the propagators and vertices always yields a traceless matrix. Therefore, they do not contribute to the grand potential. 
This also implies that the grand potential is even under $\mu \rightarrow -\mu$ transformation, as all the non-vanishing diagrams are of even orders in $\mu$.

The non-analyticities of the grand potential comes from the resummation of all IR divergent diagrams. 
We isolate these diagrams by analyzing the IR scaling of each diagram using dimensional analysis. The contribution to the grand potential from diagrams with $N=2n$ external legs scales as
\begin{equation}\label{power}
\Omega_N\sim\mu^Nk^{d+1-N}v_f^{1-N}.
\end{equation}
This indicates that diagrams with $N\geq d+1$ have infrared divergences and require resummation.

In 2D, standard calculations suggest the following structure after resummation of all IR divergent diagrams with $N=2n\geq4$ external legs
\begin{equation}
\Omega^{\text{2D}}_{\text{NA}}=\lim_{\epsilon\to0^+}\mu^3\sum_{n=2}^\infty A_{2n}\left(\frac{\mu}\epsilon\right)^{2n-3}v_f^{1-2n},
\end{equation}
where $A_{2n}$'s are the coefficients of each diagram and the infrared momentum scale $\epsilon$ approaches zero.
The resummation result (see Eq. (\ref{phi2D}) and Appendix \ref{resummation}) suggests the following non-analytical scaling function 
\begin{eqnarray}\label{phi2D}
\Omega^{\text{2D}}_{\text{NA}}\propto\frac{|\mu|^3}{v_f^2}.
\end{eqnarray}
Therefore, we can effectively treat $\epsilon$ as an IR regulator that scales as $|\mu|/v_f$.

The detailed calculation of the prefactor of this cusp  is carried out in appendix \ref{resummation} and  we present the result here
\begin{equation}
\Omega^\text{2D}=\frac{|\mu|^3}{12\pi v_f^2}+\text {analytical terms}.
\end{equation}
This structure suggests that in 2D the topological phase transitions involving Majorana fermions are 3rd order transitions with corresponding cusps in the grand potential.

The third order phase transitions in 2D $p_x+ip_y$ spinless superfluids were pointed out in a previous study. In Ref. \cite{Rombouts10}, a solvable pairing model only taking into account the pair scattering in the $({\bf k, -k})$ channel was used to analyze the phase transition. As the scatterings in non-zero total momentum channels are all absent in the model, the fluctuation effects are suppressed by construction.

As a consequence of the cusp in the grand potential, the effective compressibility $\kappa$ defined in terms of the effective chemical potential $\mu$, $\kappa=-\partial^2\Omega/\partial\mu^2$, also has a non-analytical part that can be measured experimentally.
Isolating the non-analytical part of the grand potential, we obtain the non-analytical part of compressibility in 2D,
\begin{align}
\kappa_\text{NA}^{\text{2D}}=-\frac{|\mu|}{2\pi v_f^2}.
\end{align}

In general, the diagrammatic expansion is applicable to other spatial dimensions if the transitions involve Majorana fermions. However, there is a subtle difference between even and odd dimensions as demonstrated below. 

From Eq. (\ref{power}), we conclude that all diagrams with $N=2n \geq d+1$ external legs are infrared divergent and require resummation. 
In even spatial dimensions, $d+1$ is odd and the above condition indicates that all infrared divergent diagrams must have $n \geq (d+2)/2$ and the resummation takes a form similar to the 2D case
\begin{equation}
\Omega^{\text{even}}_{\text{NA}}=\lim_{\epsilon\to 0^+}\mu^{d+1}\sum_{n=(d+2)/2}^\infty A_{2n}\left(\frac{\mu}\epsilon\right)^{2n-d-1}v_f^{1-2n},
\end{equation}
where the first term in the summation approaches infinite as $1/\epsilon$.
Like in the 2D case, one can show that the divergence can be effectively regularized by setting $\epsilon$ as an IR regulator proportional to $|\mu|/v_f$. The non-analytical part of the potential has the following form after resummation
\begin{equation}
\Omega^{\text{even}}_\text{NA}\propto\frac{|\mu|^{d+1}}{v_f^d},
\end{equation}
where the $(d+1)$th derivative is discontinuous at $\mu=0$ with different limits on two sides. So the phase transition is $(d+1)$th order.

However, in odd spatial dimensions, all diagrams with $2n$ external legs and $n\geq(d+1)/2$ have infrared divergence. Therefore, there always exists a marginally divergent diagram with $N=2n=d+1$ legs and the resummation result becomes
\begin{equation}
\begin{split}
\Omega_{\text{NA}}=&\lim_{\epsilon\to0^+}\mu^{d+1}\bigg[A_{d+1}v_f^{-d}\ln\frac{\Lambda_c}\epsilon\\
&+\sum_{n=(d+3)/2}^\infty A_{2n}\left(\frac{\mu}\epsilon\right)^{2n-d-1}v_f^{1-2n}\bigg].
\end{split}
\end{equation}
where we have isolated the marginal contribution $\ln (\Lambda_c/\epsilon)$ from the rest of the summation.
Note here the first term in the summation is divergent as $1/\epsilon^2$.
Following the same argument, the resummation suggests the non-analytical form
\begin{equation}
\Omega_\text{NA}^\text{odd}\propto\frac{\mu^{d+1}}{v_f^d}\ln\frac{|\mu|}{v_f^2}.
\end{equation}
Different from even dimensions, the $(d+1)$th derivative of $\Omega_\text{NA}$ is now weakly (logarithmically) divergent at $\mu=0$.
The phase transition again is $(d+1)$th order for odd dimensions. 

In conclusion our analysis suggests that in even dimensions, the topological phase transitions involving Majorana fermions are $(d+1)$th order transitions, with discontinuities in $(d+1)$th order derivatives.
In odd dimensions it is also $(d+1)$th order but with divergent $(d+1)$th order derivatives at the critical point. 
These signatures induced by the bulk Majorana fermions always appear along with the well-known emergence of zero-energy edge modes in topological superfluids across the transition.
Later, we compute the leading non-analytical terms quantitatively in 1D and 3D and present the result in section \ref{susy} (and Appendix \ref{resummation}).

So far we have computed the cusps via the fixed point field theory completely neglecting other terms.  Practically there are always additional interactions or perturbations potentially causing deviations from the cusp structures we have obtained. A question arises: what aspect of the cusps will be affected or not affected by which perturbations.
Below we classify and study such effects. 

Generally speaking, there are two types of perturbations to the fixed point action. The first type modifies the dispersion of free Majorana fermions; and the second type describes the interactions. 
In this section, we focus on the non-interacting case and show that the UV relevant gradient terms that alter the dispersion relation do not affect the leading order of the cusps at all (Sec. \ref{non-relativistic}). 
Then in Sec. \ref{interactions}, we analyze the effect of interactions on the cusps. Although the interaction do not change the type of non-analyticity in the grand potential, they do modified the amplitude of the cusps. 

\subsection{Robustness of the cusps: effects of gradient terms from non-relativistic dispersion}\label{non-relativistic}

The first class of perturbations is due to gradient terms of the following form,
\begin{eqnarray}\label{gradient}
\mathcal{L}_G= \frac12\overline\psi\left(-\frac{\nabla^2}2\tau_z\right)\psi.
\end{eqnarray}
It arises naturally in microscopic theories of superfluids reflecting that the underlying physics is non-relativistic and the relativistic feature of the effective theory only emerges in the infrared limit. 
Once the system is in the relatively high energy scales, the non-relativistic nature of underlying fermions always re-emerges, leading to deviations from the relativistic physics.
Here we show that the non-relativistic correction does not enter the leading order of the non-analytical structure at all, not  even in a quantitative way. 
It only modifies the regular terms or higher order non-analyticities which are not the concern of the current study. 
For instance, in 2D not only the critical indices are unaffected as one expects, but even the leading cusp structure of the grand potential is immune of these non-relativistic effects.

\begin{figure}
\includegraphics[width=\columnwidth]{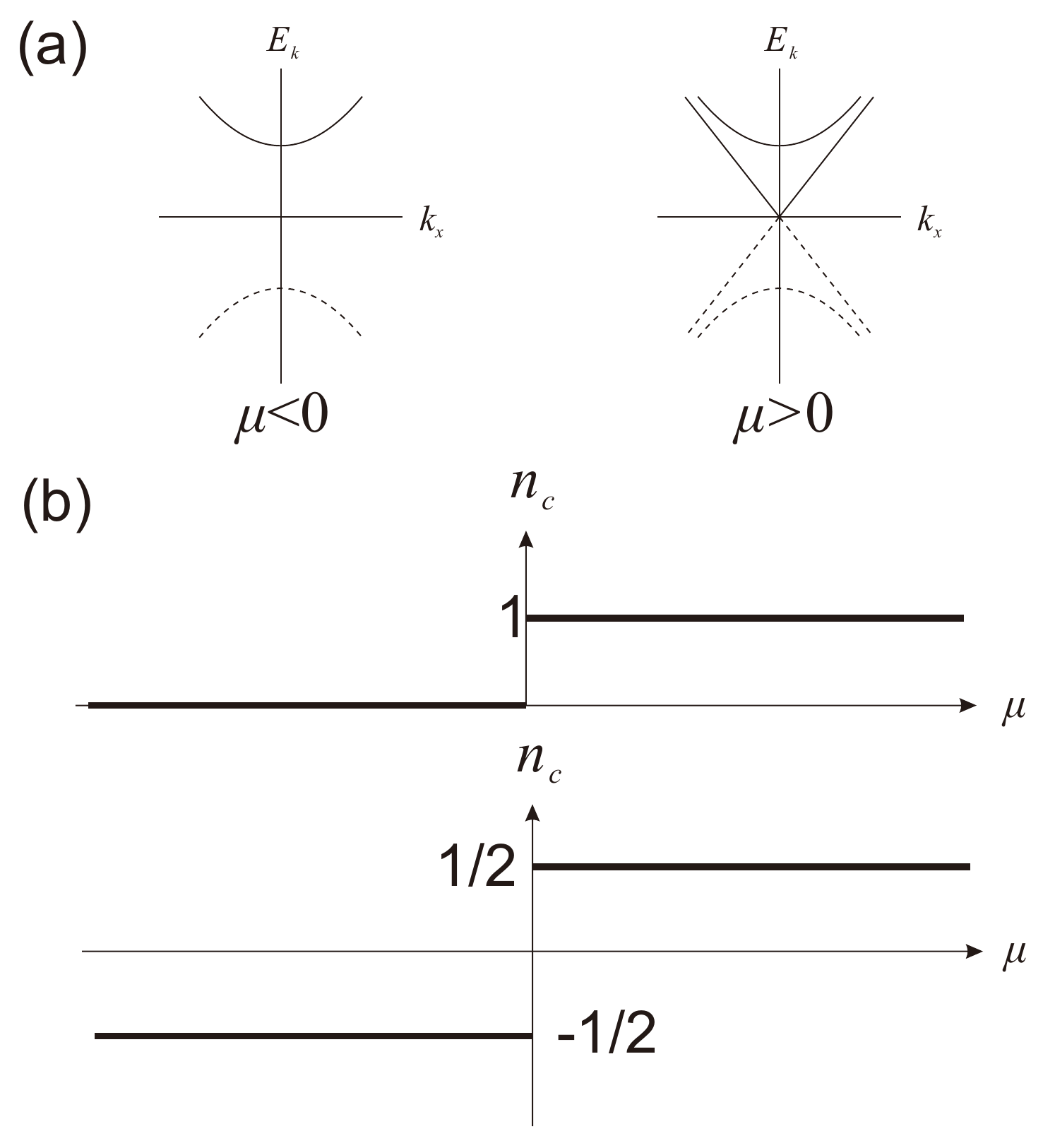}
\caption{Topological phase transitions driven by the effective chemical potential $\mu$. We demonstrate the change of topology in a spinless system using Chern number $n_c$ and chiral edge modes in a 2D strip geometry. (a) For $\mu<0$ the superfluid is topologically trivial with no edge mode; while for $\mu>0$ the superfluid is topologically non-trivial with  chiral edge modes. 
(b) Chern number $n_c$ defines the topology of a state and generally depends on the UV behaviour of the theory. Here we show the instability of the Chern number with respect to infrared irrelevant operators in an extended theory with the UV cut-off $\Lambda_c\to\infty$. The top is for an extended theory with an additional irrelevant gradient term in Eq. (\ref{gradient}); the bottom is for the fixed point theory without the gradient term. 
Note that the edge mode physics remains largely unaffected by such a gradient.
}\label{topology}
\end{figure}

We treat the gradient term from the non-relativistic dispersion [Eq. (\ref{gradient})] as a separate vertex and perform a similar dimensional analysis on the one-loop diagrams involving both the gradient term and effective chemical potential [Fig. \ref{diagram}(b)]. 
Again as the total number of vertices is always equal to the number of propagators in one-loop diagrams, all diagrams with odd number of fermion propagators vanish due to the traceless $\tau_z$ matrix in each vertex; and the total number of vertices is also even.

For a diagram with $N$ effective chemical potential and $N'$ gradient terms with $N+N'$ even, its contribution to the grand potential scales as
\begin{equation}
\Omega_{N,N'}\sim\mu^Nk^{N'-N+d+1}v_f^{N+N'-1}
\end{equation}
in the IR regime.
As we can see, for the same order of $\mu$ (i.e. fixed $N$), the more gradient terms (i.e. the larger $N'$ is), the less singular $\Omega_{N,N'}$ is in the infrared limit. 
For even $N$, diagrams without the gradient term give the leading infrared singularity.
For odd $N$, all the diagrams without gradient terms vanish, and diagrams with one gradient term give the leading infrared singularity. 
For example, in 2D not only diagrams with $N'=0$ and $N\geq4$ are singular, diagrams with $N'=1$ and $N\geq5$ also have infrared divergences and require resummation. 
For $N'=0$, the resummation gives the same non-analytical term $|\mu|^3$ as the previous section without the gradient term; for $N'=1$, similar resummation procedure gives the non-analytical term of higher order of $\mu$. 

In conclusion, the gradient term from non-relativistic dispersion does not alter the leading non-analytical structure in the grand potential at all. However, it does generate higher order non-analytical terms.

Before ending this part of discussions, we comment on the other seemingly surprising aspect of  the gradient terms.
Although the gradient term is infrared irrelevant from the standard scaling point of view and does not contribute to the leading non-analytical structures at all, it does change the UV  physics, and hence can result in the change of topology. 
This is because topology is a global property depending on the entire energy spectrum including the UV part, while the leading order of the cusps only depend on the low energy physics. 

This can be clearly illustrated by considering the deviation from the fixed point action due to this gradient term. For the convenience of the discussions on topology and edge modes, instead of using the fixed point Lagrangian in Eq. (\ref{fpl}), we use the fixed point (FP) Hamiltonian density
\begin{eqnarray}
\mathcal{H}_{\text{FP}}= \frac12\psi^\dagger \left[\vec{\tau} \cdot (-i\nabla) -  \mu\tau_z \right]\psi
\label{2dFP}
\end{eqnarray}
and extend the UV cut-off to infinity, i.e. $\Lambda_c\to\infty$, so that we can define its topology.
The Chern number associated with this extended model is $n_c(\mu > 0)=1/2$ and $n_c(\mu <0)=-1/2$.
$\delta n_c=n_c(\mu>0)-n_c(\mu<0)=1$ as one varies the effective chemical potential from a negative value to a positive value, so that the interface of the two states with opposite signs in effective chemical potentials always has a Majorana mode present, a well known fact. We can further use the extended model to define a physical vacuum with no fermions; one can achieve this by taking the limit $\mu_{vac}\to -\infty$.
A 2D disk then supports a Majorana edge mode when $\mu >0$ and no edge modes otherwise.

Next, we consider a theory defined by $\mathcal{H}_{\text{FP}} +\mathcal{H}_G$ by including the gradient term in Eq. (\ref{gradient}), where we rewrite the gradient term using the Hamiltonian density
\begin{eqnarray}\label{gradient}
\mathcal{H}_G= \frac12\psi^\dagger\left(-\frac{\nabla^2}2\tau_z\right)\psi.
\end{eqnarray}
The Chern number $n_c$ is shifted by one-half. Now,  $n_c(\mu<0) \rightarrow  0$ and $n_c(\mu>0) \rightarrow 1$.
However, the difference between the Chern numbers on the two sides of the transition $\delta n_c$ remains invariant with respect to the addition of the gradient term (Fig. \ref{topology}). 
One can further show that the existence or absence of edge modes will also remain unaffected when including such a gradient term, mainly because the edge Majorana mode itself, just like the leading order behaviour of the cusps, only depends on the infrared limit of the theory when the gradient term $\mathcal H_G$ becomes irrelevant. Only the spatial structure of the Majorana modes will be slightly perturbed and deformed due to $\mathcal{H}_G$ in the limit of interest.  
Therefore, one can still describe the topological phase transition by the effective theory neglecting the gradient term from non-relativistic dispersion.

The shift of the Chern number by one-half unit can be simply attributed to a change of the Hamiltonian manifold when the gradient term is added.
The Hamiltonian manifold without the gradient term is equivalent to one-half of a two-sphere $S^2$ while it turns into a two-sphere $S^2$ when the gradient term is present.
In the later case, the Hamiltonian defined in $k$ space effectively results in mappings from $S^2$ to $S^2$ and it can be topologically non-trivial and is classified by the second homotopy
group of $S^2$, $\pi_2(S^2)=Z$. Only $\mu >0$ fall into a non-trivial class while $\mu <0$ corresponds to a trivial mapping.
Formally speaking, since at each ${\bf k}$ point eigenmodes of the Hamiltonian $\mathcal H_{FP}+\mathcal H_G$ are non-degenerate, there can be a one to one mapping between the classical manifold of the Hamiltonian characterized by its homotopy groups, and the state vector subspace (with either positive or negative energies) characterized by Chern-numbers. Physically, this indicates that the topology of many-body ground state  shall be unique for a given equivalent class
of the Hamiltonian. Classification of the Hamiltonian manifold in this case provides a convenient way to classify superfluids.

\section{Robust structures of cusps: renormalization effects of interactions}\label{interactions}

\begin{figure}
\includegraphics[width=\columnwidth]{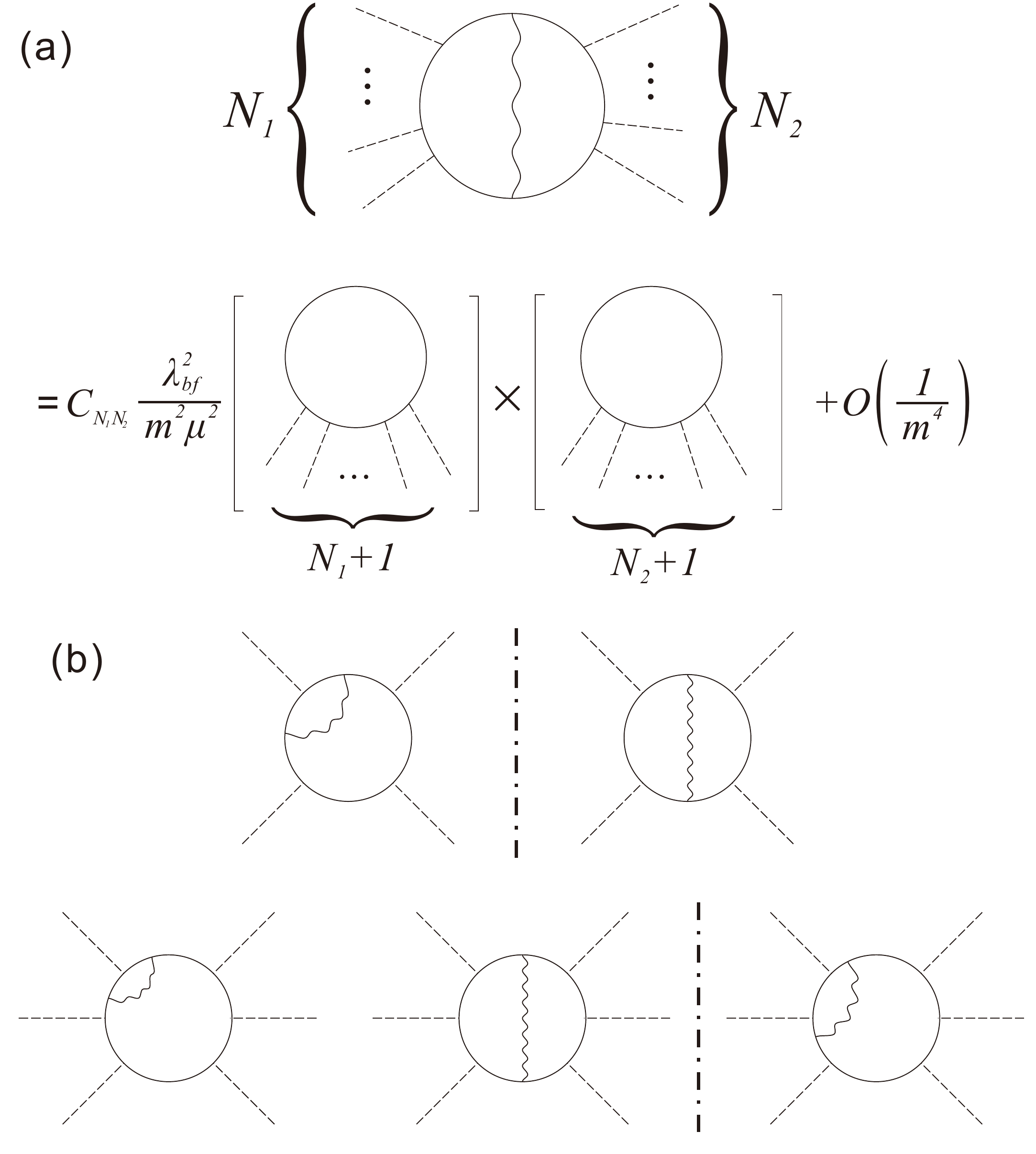}
\caption{
We consider the effect of Yukawa interactions in Eq. (\ref{FPaction}) by including multi-loop diagrams. The solid lines represent gapless free relativistic fermion propagators, the dashed lines represent effective chemical potential $\mu$ which is treated as an external field and the wavy lines represent the free but massive boson propagators. (a) In the $1/m$ expansion, a two-loop diagram can be approximated by a product of two one-loop diagrams, as is shown here. $C_{N_1,N_2}$ is the coefficient of each diagram, particularly $C_{1,N-1}=-N$ (notice each fermion loop contains a factor of $-1$). 
(b) Examples of infrared divergent two-loop diagrams with four and six legs. All the two-loop diagrams can be organized by the number of external legs. Diagrams have non-vanishing contributions to the order of $1/m^2$ if and only if both $N_1$ and $N_2$ are odd. The vanishing and non-vanishing diagrams are separated by the thick dash-dotted lines.
}\label{diagram2}
\end{figure}

In the previous section, we have discussed the non-analytical cusps in the limit of Majorana fermion fixed point where Majorana fermions propagate freely without interactions.
We have shown that these non-analytical cusps are very robust, uniquely associated with the infrared physics described in the effective theory, and its leading term cannot be altered by simply varying the UV physics, such as adding a UV relevant gradient term discussed above.
Now we turn to the interesting effect of the interactions between Majorana fermions and the fluctuating pairing fields on the cusps. 
The limit of interacting Majorana fermions is more generic, naturally emerging at phase transitions due to the coupling to various fluctuations. 
The question we are answering here is whether and how the cusps can be used to detect those renormalization effects of the interactions between Majorana fermions and the fluctuations. 

We show that the interactions do affect the cusp structures from the fixed point theory. The effect is closely related to the renormalization of Majorana fermion fields when interacting with the massive fluctuating bosonic fields. 
In general, the renormalized Majorana fields can receive vertex corrections, have renormalized field strengths and renormalized velocity. In the large $m$ limit, we can isolate the effect of $\mu$-vertex (effective chemical potential) renormalization from the higher order field strength and velocity renormalization effects.
Although the order of phase transition is not affected by these effects, the amplitude of the non-analytical cusps can be modified; most importantly the cusp measurement yield information of how the Majorana fermion fields interact with fluctuating bosonic fields. 

Following the discussions in Sec. \ref{fixedpointH}, we focus on the Yukawa interaction $\lambda_{bf}$ which represents the least irrelevant interaction compared with other interactions for our purpose.
It has the least negative scaling dimension among all irrelevant operators and is also the leading term in the $1/m$ expansion we apply below.

Corrections to the grand potential come from multi-loop diagrams with internal bosonic propagators. These diagrams can be conveniently organized in powers of $1/m^2$. 
In the following, we focus on the corrections to the leading order of $1/m^2$. This involves all two-loop diagrams with one internal boson propagator (Fig. \ref{diagram2}).
 One can obtain these diagrams by adding a pair of Yukawa vertices $\lambda_{bf}\tau_z$ and an internal boson propagator to one-loop diagrams of the fixed-point action. This increases the number of fermion propagators by two compared with the corresponding one-loop diagram. 
Following the previous section, since the Yukawa interaction also involves the traceless $\tau_z$ matrix and the total number of vertices still equal to the number of fermion propagators in these two-loop diagrams, only diagrams with even number of fermion propagators contribute to the grand potential, and all the two-loop diagrams are even orders of $\mu$.

To compute the effect of interactions, we first demonstrate that one can approximate a two-loop diagram by a product of two one-loop diagrams of the fixed point action multiplied by a proper prefactor [see Fig. \ref{diagram2}(a)].
Consider a two-loop diagram with $N_1$ external legs on one side and $N_2$ legs on the other side, separated by a bosonic propagator. 
Since $G_B^{(0)}{}^{-1}\approx-m^2$ is independent of frequency or momentum, the two integrations in the diagram are separable in the leading order of $1/m^2$. 
Each separate integration effectively describes a one-loop diagram with $N_1+1$ and $N_2+1$ fermion propagators, respectively. Thus, such diagrams, to the leading order of $1/m^2$, can be approximated by a product of two one-loop diagrams with a prefactor $C_{N_1,N_2}\lambda_{bf}^2/(m^2\mu^2)$.
$C_{N_1,N_2}$ is a numerical factor associated with the multiplicity of each diagram. In particular, $C_{1,N-1}=-N$. The negative sign comes from the fact that each fermion loop contains a factor of $-1$. 
Higher order corrections can be included by expanding the boson Green's function [see Eq. (\ref{boson})], but these terms are at least order of $1/m^4$ and do not concern us here.
The above approximation significantly simplifies our analyses.

Next, we classify all two-loop diagrams by the order of $\mu$ and their IR behaviour. We demonstrate that only a fraction of these diagrams contribute to the leading non-analytical structure of the grand potential.
We first notice that some of the two-loop diagrams vanish in the above approximation. Since we can approximate each two-loop diagram with a product of two one-loop diagrams, each one-loop diagram has to have even number of fermion propagators for the result not to vanish. Therefore, we only need to consider diagrams in which both $N_1$ and $N_2$ are odd. 
Furthermore, some two-loop diagrams are less singular than others in the IR limit. 
For diagrams of the same order of $\mu$, some of them are less divergent than the one-loop diagram with the same number of $\mu$-legs, and thus only contribute to higher order non-analytical terms.
There is only one non-vanishing four-leg diagram. For diagrams with six legs or more, there are more than one non-vanishing diagrams. However, not all of them contribute to the leading non-analytical term.
We use the six-leg diagrams as an example. There are two non-vanishing six-leg diagrams [Fig. \ref{diagram2}(b)]. Since the one-loop diagrams diverges as $\mu^N/\epsilon^{N-3}$ if and only if $N\geq4$ in 2D, the first  diagram with $N_1=1, N_2=5$ is divergent as $\mu^6/\epsilon^3$, same as the one-loop six leg diagram; while the second one with $N_1=N_2=3$ diverges as $\mu^6/\epsilon^2$. 
As a result, the second diagram only contributes to non-analyticities of higher order in $\mu$. Similarly, for $N$-leg two-loop diagrams, only the diagrams with $N_1=1, N_2=N-1$ and $C_{1,N-1}=-N$ contribute to the non-analytical $|\mu|^3$ structure in 2D.

Adding these two-loop diagrams to the IR singular one-loop diagrams, we have
\begin{equation}
\begin{split}
&\sum_{n=1}^\infty A_{2n}\epsilon^{3-2n}v_f^{1-2n}(\mu^{2n}+2n\mu^{2n-1}\delta\mu)\\
=&(\mu+\delta\mu)^3\sum_{n=1}^\infty A_{2n}\left(\frac{\mu+\delta\mu}\epsilon\right)^{2n-3}v_f^{1-2n}+O(\delta\mu^2),
\end{split}
\end{equation}
where $\delta\mu=-\lambda_{bf}^2\Lambda_c\mu/(2\pi^2m^2v_f)$. 
This indicates that the leading order effect of the interactions between Majorana fermions and the fluctuations is to renormalize the effective chemical potential from $\mu$ to $\mu+\delta\mu$. 
As a result, if one expresses the grand potential of the interacting system using the renormalized effective chemical potential, the form of the cusp remains the same to the leading order of $1/m^2$,
\begin{equation}
\Omega^{\text{2D}}_{\text{NA}}=\frac{|\mu+\delta\mu|^3}{12\pi v_f^2}+O\left(\frac1{m^4}\right).
\end{equation}
This indicates that the cusp structure is robust against interactions due to fluctuations.
Alternatively, we can also express this resultant grand potential in terms of the original effective chemical potential $\mu$
by keeping the terms to order $1/m^2$,
\begin{equation}
\Omega^{\text{2D}}_{\text{NA}}=\frac1{12\pi v_f^2}\left(1-\frac{3\lambda_{bf}^2\Lambda_c}{2\pi^2m^2v_f}\right)|\mu|^3+O\left(\frac1{m^4}\right),
\end{equation}
and the leading non-analytical term of the effective compressibility is
\begin{align}
\kappa_\text{NA}^{\text{2D}}=-\frac{|\mu|}{2\pi v_f^2}\left(1-\frac{3\lambda_{bf}^2\Lambda}{2\pi^2m^2v_f}\right).
\end{align}
Note that this approximation does not fully describe the $O(1/m^4)$ correction.

Before closing our discussions on 2D isotropic superfluids and superconductors,  let us point out that the above discussions can be straightforwardly generalized to time reversal symmetric topological 
superfluids or superconductors discussed in the literature \cite{Qi09, Sato09}. The time reversal symmetric states can be constructed by putting two copies of superfluids, $p_x\pm ip_y$ together with $\uparrow\uparrow$ and
$\downarrow\downarrow$ spins respectively.
This  can be shown to be equivalent to the projection of a three dimensional Balian-Werthamer state \cite{Leggett75} onto a two dimensional plane of $k_z=0$, with $\Delta({\bf k})=i\sigma_y \vec\sigma \cdot {\bf d}({\bf k})$, $\vec\sigma=(\sigma_x,\sigma_y)$ and ${\bf d}=k_x {\bf e}_x+k_y {\bf e}_y$. $\sigma_{x,y,z}$ are Pauli matrices in spin space.
These two copies effectively form Kramer doublets which are connected by the time reversal
transformation. 

The fixed point Hamiltonian takes a simple form of 
\begin{equation}
\mathcal{H}^{TRS}_{\text{FP}}=\frac12\psi^\dagger \left[\tau_x \otimes \sigma_z  (-i\partial_x) -\tau_y \otimes I (-i\partial_y) - \tau_z \otimes I  \mu \right]\psi.
\end{equation}
where the superscript {\em TRS} indicates the time reversal symmetry.  
The spinful Nambu fermion is defined by $\psi=(c_{\uparrow}({\bf r}), c_{\downarrow}({\bf r}), c_{\uparrow}^\dagger({\bf r}), c_{\downarrow}^\dagger({\bf r}))^T$ .

We construct the corresponding effective theory near the fixed point by generalizing the interactions in the single-copy theory above. 
\begin{align}
&\mathcal L= \mathcal L_{\phi_{+,-}} +\mathcal L_{\psi} +\mathcal L_{\psi\phi_{+,-}}; \nonumber \\
& \mathcal L_{\phi_{+,-}}=-\frac12\sum_{i=+,-}\phi_i (\partial^2_\tau+v_b^2\nabla^2-m^2)\phi_i+\lambda_b \phi_i^3+g_b \phi_i^4; \nonumber\\
&\mathcal L_{\psi}=\frac12\overline\psi \left[\partial_\tau+\tau_x \otimes \sigma_z  (-i\partial_x) -\tau_y \otimes I (-i\partial_y) - \tau_z \otimes I  \mu \right]\psi; \nonumber \\
&\mathcal L_{\psi\phi_{+,-}} = \frac12\lambda^{e}_{bf} (\phi_++\phi_-) \overline\psi\tau_z\otimes I\psi\nonumber\\
&\qquad\qquad+\frac12\lambda^{o}_{bf} (\phi_+-\phi_-) \overline \psi \tau_z \otimes \sigma_z  \psi.
\end{align}
Here $\phi_{+, -}$ represent the pairing amplitude fluctuation fields in $p_x\pm ip_y$ channels respectively. $\lambda^{e,o}$ are for interactions in symmetric and antisymmetric pairing fluctuation channels.
The interaction is invariant under the time reversal symmetry transformation. 

There should be two Majorana edge modes related to each other by the  time reversal symmetry in the topological phase but no edge modes in the trivial phase.
For a system of strip geometry with upper and lower edges located at $y=0$ and $y=y_0$, the edge mode Hamiltonian is 
\begin{align}
&\mathcal H^\text{edge}=(-1)^{1+P_{u,l}}\frac12\psi_{u,l}^\dagger [v_f\sigma_z(-i\partial_x)]\psi_{u,l},\nonumber\\
&\psi_{u,l}(x)=\frac12(1\pm\tau_z)e^{i(\pi/4)\tau_y}\psi(x).
\end{align}
where $P_u=1$ and $P_l=0$ are the eigenvalues of chirality projection operators $\frac12(1+\tau_z)$ corresponding to the upper and lower edges, respectively.
The edge modes are in the class of symmetry protected topological states and are robust to non-magnetic disorders \cite{Qi09,Chen13}.
The cusps have the same feature as discussed above as long as superfluids are isotropic.

\begin{figure}
\includegraphics[width=\columnwidth]{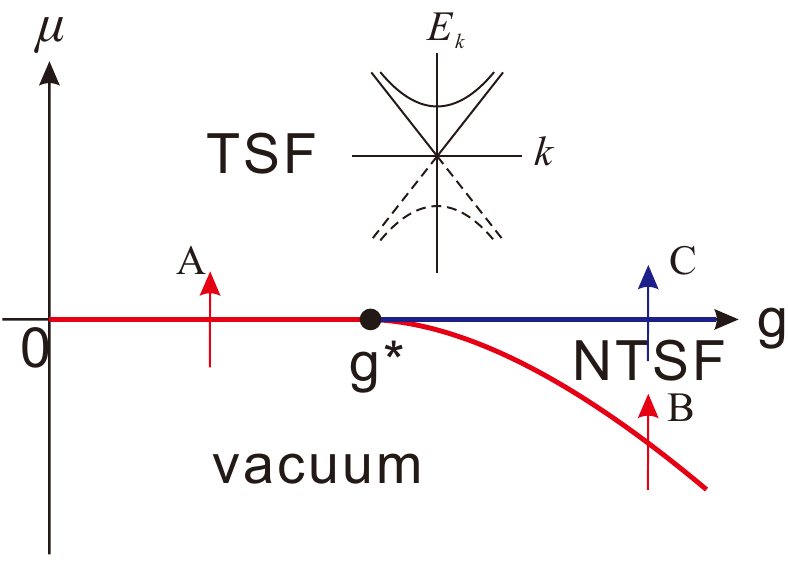}
\caption{Schematic of phases of interacting fermions with $p$-wave pairing. $\mu$ is the effective chemical potential of fermions and $g$ is the $p$-wave coupling constant in the underlying microscopic model \cite{ftRev}. For weak interactions ($g<g^*$), the system undergoes a phase transition between the topological superfluid (TSF) phase and vacuum at $\mu=0$ (labeled by arrow A). For strong interactions ($g>g^*$), the vacuum to superfluid phase transition (labeled by arrow B) is expected to occur at negative effective chemical potential, $\mu=\mu_c(g)$. For $\mu_c(g)<\mu<0$, the system is in the non-topological superfluid (NTSF) phase. A topological phase transition occurs at $\mu=0$ (labeled by C) for $g>g^*$. The inset shows the gapped bulk dispersion and gapless boundary modes in the TFS phase of a system with strip geometry. \label{Phase}}
\end{figure}

To conclude this section, we present the schematics of phases of the $p$-wave interacting ferimons in Fig. \ref{Phase}. Here we assume no other instabilities than pairing, and continuous phase transitions. In the inset, we show the dispersion of the edge modes in the strip geometry. 
Note that left- and right-moving modes are located on opposite edges for the case of chiral superfluids; while in TRS superfluids, there are a pair of Kramer doublet edge modes on each edge. 
In the disk geometry, there is only one edge mode for chiral superfluids, while two modes for TRS superfluids which are Kramer doublets.  The gapless modes are either topologically protected in chiral superfluids, or symmetry protected in time reversal symmetric superfluids. In both cases, the phase transition in 2D is of the third order.

\section{Relation to supersymmetry models and Cusps in 1D and 3D} \label{susy}

To discuss the relation to the $\mathcal{N}=1$ minimum supersymmetry model and the topological phase transitions in other spatial dimensions, it is beneficial to cast the effective theory in a standard Majorana fermion representation, where two-component Majorana fermions explicitly  couple to a single real scalar field. 
One can then straightforwardly extend the cusp analysis to 1D and 3D as long as those topological phase transitions are also in the universality class of free Majorana fermions, disregarding the microscopic origin.

We first introduce the field operators in the Majorana representation
\begin{eqnarray}
&& \phi ({\bf r}) =\sigma({\bf r}),\; \Psi ({\bf r}) =(\gamma_1({\bf r}), \gamma_2({\bf r}))^T; \nonumber \\
&& \gamma_1 ({\bf r}) =\frac{1}{\sqrt{2}}(c({\bf r})+c^\dagger({\bf r})), \nonumber \\
&& \gamma_2 ({\bf r})=\frac{1}{i\sqrt{2}}(c({\bf r})-c^\dagger({\bf r})).
\end{eqnarray}
$\gamma_1({\bf r})$ and $\gamma_2({\bf r})$ are real Majorana fermion operators that obeys
\begin{align}
\gamma_\alpha^\dagger({\bf r})=\gamma_\alpha({\bf r}), && \{\gamma_\alpha({\bf r}),\gamma_\beta({\bf r'})\}=\delta_{\alpha,\beta}\delta({\bf r}-{\bf r'})
\end{align}
with $\alpha,\beta=1,2$. 
In this representation, the two-component fermion operators $\Psi({\bf r})$ are real fermion operators satisfying
\begin{align}
\Psi^\dagger({\bf r})=\Psi^T({\bf r}), && \{\Psi({\bf r}),\Psi({\bf r'})\}=\delta({\bf r-r'}).
\end{align}
Notice that in the momentum space, the corresponding fields have the following properties
\begin{align}
\Psi_{\bf k}=(\gamma_{1,{\bf k}},\gamma_{2,{\bf k}})^T, && \{\Psi_{\bf k},\Psi_{\bf k'}\}=\delta_{\bf k,-k'}.
\end{align}

Using the Majorana representation, the effective field theory for the 2D $p_x-ip_y$ superfluid can be rewritten as 
\begin{eqnarray}\label{LagrangianM}
&&\mathcal L= \mathcal L_{\sigma} +\mathcal L_{\Psi} +\mathcal L_{\Psi\sigma} +\mathcal L^{SSB}_{\{\Psi,\sigma\}};
\nonumber \\
&& 
\mathcal L_{\sigma}=-\frac12\sigma(\partial^2_\tau+v_b^2\nabla^2-m^2)\sigma+
g_b\sigma^4; \nonumber\\
&&\mathcal L_{\Psi}=
\frac12\overline\Psi [\partial_\tau-i v_f (\tau_z {\partial}_x -{\tau}_x {\partial}_y)+ \mu\tau_y] \Psi; \nonumber \\
&&\mathcal L_{\Psi\sigma} = - g_{bf} \sigma^2 \overline \Psi\tau_y \Psi; \nonumber\\
&& \mathcal L^{SSB}_{\{\Psi,\sigma\}} =-\lambda_{bf}\sigma \overline \Psi\tau_y\Psi+\lambda_b\sigma^3.
\label{2dL}
\end{eqnarray}
Here $\overline\Psi({\bf r},\tau)=\Psi^T({\bf r},\tau)$.

This effective action at first sight appears to resemble the $\mathcal{N}=1$ supersymmetric Yukawa-Gross-Neveu model. The latter model in two spatial dimension has been recently studied\cite{Fei17} and is suggested to flow into a supersymmetric conformal fixed point in $(2+1)$D. 
However, there are two stark distinctions. One is that the real scalar field is massive representing a {\em disorder fixed point} from the fluctuation point of view.
Equally importantly, because the transition occurs at a symmetry breaking state, the model generically contains a cubic term of $\sigma^3$, which is more relevant than the standard $Z_2$ symmetry preserving quartic term of $\sigma^4$. 
These two terms represent relevant operators in the massless limit, so unless both the mass term and cubic term are further fine tuned to vanish, the dynamics near the topological transitions
considered here should be distinct from that of the $\mathcal{N}=1$ supersymmetry model and in general does not exhibit emergent supersymmetries. 
By the same token, its projection onto 1D (discussed below) and the corresponding phase transitions also do not exhibit the dynamics of the $\mathcal{N}=1$ minimum supersymmetry model in $(1+1)$D.
  
In addition, the cubic term seems to suggest the possibility of a discontinuous phase transition near this critical point.  This possibility is also hinted in a recent energetic analysis of the stability of $p$-wave superfluid induced by short range resonance interactions \cite{Jiang18}.
More efforts are needed to understand the relations between different considerations. For example, under which condition a discontinuous transition occurs, and what is the dependence on the microscopic details.
Here we only consider the continuous transition.

In the previous sections, we mainly focus on 2D Majorana fermions.
Using the Majorana representation, we can easily generalize our results to one and three dimensions. If a topological transition is characterized by Majorana Fermi gases,  generically the fixed point Lagrangian can always be written as
\begin{eqnarray}
&& \mathcal L_{\Psi}=
\frac12\overline{\Psi} [\partial_\tau+\mathcal{H}_{FP}^{(d)}] \Psi, \nonumber \\
&& \mathcal{H}_{FP}^{(d)} =\left\{
\begin{array}{cc}
-i v_f \tau_z \partial_ x  + \mu\tau_y,  & \mbox{$d=1$;}  \\
-i v_f \vec \Gamma\cdot\nabla + \mu\Gamma_0, & \mbox{$d=3$}.
\end{array}
\right.
\end{eqnarray}
Here $\tau_{x,y,z}$ are Pauli matrices in the $\gamma_1,\gamma_2$ subspace, $\nabla$ is the 3D gradient operator, and  $\vec\Gamma=(\Gamma_x,\Gamma_y,\Gamma_z)$ and $\Gamma_0$ are Majorana matrices of the form
\begin{eqnarray}
&& \Gamma_x=\tau_z \otimes \sigma_z,\quad
\Gamma_y=\tau_x \otimes I,
\nonumber \\
&& \Gamma_z=-\tau_z \otimes \sigma_x,\quad
\Gamma_0=\tau_y\otimes I,
\end{eqnarray}
with $\sigma_{x,y,z}$ being the standard Pauli matrices in spin space. In 1D, $\Psi$ is the same two-component spinless Majorana field; $\mathcal{L}_\Psi$ for 1D can be conveniently obtained by projecting the two-dimensional $\mathcal{L}_{\Psi}$ in Eq. (\ref{2dL}) onto the $k_y=0$ plane.

In 3D we have generalized $\Psi$ to a four-component spinful Majorana field to include spins, and $\Psi$ is defined as
\begin{eqnarray}
\Psi({\bf r},\tau)=(\gamma_{1,\uparrow}, \gamma_{1,\downarrow}, \gamma_{2,\uparrow}, \gamma_{2,\downarrow})^T
\end{eqnarray}
in real space.
The particular representation of $\Gamma$ matrices has been chosen for  isotropic superfluids of Balian-Werthamer (BW) type \cite{Leggett75} with the spin quantization axis $\bf{d}$ locked along the direction of momenta.
 The BW state is commonly identified with $^{3}$He-B phase with a further rotation of $\bf{d}$ by a magic angle\cite{Leggett75}.
The singularities obtained below, however, are robust, independent of the particular representation we employ to facilitate discussions, applicable to all {\em isotropic} superfluids.

 The distinction between $\mu > 0$ and $\mu <0$ superfluids in 3D is also hinted by the Hamiltonian manifold.
Just like the 2D case, we can examine the topology of the classical Hamiltonian manifold $\mathcal H_{FP}^{(d=3)}$ to infer the topology of underlying superfluids. 
To facilitate discussions, we can further add an infrared irrelevant gradient term $\mathcal H_G$ to $H_{FP}^{(d=3)}$ without affecting the infrared physics. 
The Hamiltonian inferred by $\Gamma$ matrices then forms a matrix representation of $S^3$ manifold.
The Hamiltonian effectively defines a mapping from $S^3$, the 3D momentum space with $|{\bf k}|\to\infty$ identified, to $S^3$ of the Hamiltonian manifold. Topological properties of the mappings are specified by the third
homotopy group of the classical Hamiltonian manifold $S^3$, i.e. $\pi_3(S^3)=Z$. 
One can show that the Hamiltonian with
$\mu > 0$ represents a non-trivial mapping; while with $\mu <0 $ again a trivial one implying different topologies of superfluids when the chemical changes sign.

Indeed, one can further show explicitly that the interface between such two superfluids with effective chemical potentials of opposite signs must support {\em gapless} Majorana surface states, just like in 2D, as a result of the change of topology. 
These surface states can be described by a surface Hamiltonian for interfaces perpendicular to the $y$-direction 
\begin{align}
&\mathcal H^\text{surface}=(-1)^{1+P_{u,l}}\frac12\Psi_{u,l}^\dagger[\vec\sigma\cdot(-i\nabla\times  {\bf e_y})]\Psi_{u,l},\nonumber\\
&\Psi_{u,l}(x,z)=\frac12(1\pm\tau_z)\Psi(x,z),
\end{align}
where $\vec\sigma=(\sigma_x,\sigma_y,\sigma_z)$. The subscripts $u,l$ correspond to upper and lower surface states. $P_u=1$ and $P_l=0$ are the eigenvalues of $(1+\tau_z)/2$ for $\Psi_{u,l}$ respectively.

The superfluids with $\mu>0$ hence have the helical surface states with spin and momentum locked at a right angle.
Note that the gradient term $\mathcal H_{G}$ again plays little role in surface states as they are infrared irrelevant.
As expected, superfluids with positive effective chemical potentials are topologically distinct  from the ones with negative effective chemical potentials, and there shall be a phase transition between them.

Note that the Hamiltonian $\mathcal H_{FP}^{(d=3)}$ is explicitly time-reversal invariant. Therefore, the Chern number characterization cannot be directly applied to the discussions on the superfluid topology. Instead, the time reversal symmetry implies an intimate connection to $Z_2$ topological insulators.
This was pointed out and discussed in details in Ref. \cite{Qi09, Sato09,Schnyder08,Roy08,Kitaev09}. We refer the reader to these references for more detailed discussions on the consequences of time reversal symmetry in 3D topological superfluids. 
However, the general classification of time reversal symmetric superfluids with $Z_2$ indices, formally speaking, relies on the symmetries of the Brillouin zone; this is beyond the scope of the effective theory in the continuous limit which we are discussing. 
From a more physical perspective, $Z_2$ invariants effectively specify odd or even number of helical states on each surface and hence distinguish topological versus non-topological superfluids. 
The homotopy groups of a Brillouin zone or torus have also been previously employed to classify time-reversal invariant topological insulators in Ref. \cite{Moore07}.

As noted before, the 2D cusps we obtained in Sec. \ref{cusps} (up to a prefactor) are directly applicable to time reversal symmetric superfluids in 2D as the infrared physics of isotropic superfluids is robust and independent of time reversal symmetry. By the same token, below we will focus on the general energetics and robust cusps in 3D isotropic superfluids by applying the above theory as an infrared effective description.

The cusps associated with the fixed point theory can be studied in a similar way and the results are listed below. As stated in Sec. \ref{cusps}, in odd spatial dimensions, the leading non-analyticity contains a logarithmic divergence. The resummation of IR divergent one-loop diagrams yields the following non-analytical grand potential in 3D (see Appendix \ref{resummation} for details)
\begin{eqnarray}
\Omega^{\text{3D}}=-\frac{\mu^4}{16\pi^2v_f^3}\ln{\frac{|\mu|}{v_f^2}}+\mbox{analytical terms}
\end{eqnarray}
indicating a fourth order phase transition.
Isolating the non-analytical contributions, we obtain the non-analytical part of the effective compressibility in 3D,
\begin{align}
\kappa^\text{3D}_\text{NA}=\frac{3\mu^2}{4\pi^2v_f^3}\ln{\frac{|\mu|}{v_f^2}}.
\end{align}

Just like in 2D, the interactions with massive fluctuation fields via the Yukawa term renormalize the cusp structure in 3D. Performing very similar calculations, we obtain the results listed below.
The grand potential is
\begin{equation}
\Omega^{\text{3D}}_{\text{NA}}=-\frac{\mu^4}{16\pi^2v_f^3}\left(1-\frac{\lambda_{bf,\text{3D}}^2\Lambda_c^2}{2\pi^2m^2v_f}\right)\ln{\frac{|\mu|}{v_f^2}}+O\left(\frac1{m^4}\right),
\end{equation}
and the leading non-analytical term in the effective compressibility is
\begin{equation}
\kappa^\text{3D}_\text{NA}=\frac{3\mu^2}{4\pi^2v_f^3}\left(1-\frac{\lambda_{bf,\text{3D}}^2\Lambda_c^2}{2\pi^2m^2v_f}\right)\ln{\frac{|\mu|}{v_f^2}},
\end{equation}
where $\lambda_{bf,\text{3D}}$ is the effective coupling constant in 3D.

Finally, we turn to the 1D case. Using the fixed point Hamiltonian of Majorana fermions, one finds that  

\begin{eqnarray}
\Omega^{\text{1D}}=+\frac{\mu^2}{4\pi v_f}\ln{\frac{|\mu|}{v_f^2}}+\mbox{analytical terms}.
\end{eqnarray}
Accordingly,
\begin{align}
 \kappa^\text{1D}_\text{NA}=-\frac{1}{2\pi v_f}\ln{\frac{|\mu|}{v_f^2}},
\end{align}
implying a second order phase transition.

Different from higher dimensions, the Yukawa interaction operator $\lambda_{bf}$ is marginal in 1D, so it can potentially lead to a different marginal divergent behavior. We have performed a summation over the most divergent diagrams that are of order of $(\ln |\mu|)/m^2$ and proportional to $\lambda_{bf}^2$. These diagrams always involve a boson propagator going around a single external $\mu$-leg leading to the renormalization of $\mu$ [See Fig. \ref{diagram}(b)].
Among the complete set of diagrams of order of $\lambda_{bf}^2$, we neglect all other sub-leading diagrams of order of $1/m^2$ without the logarithmic factors; this class of diagrams involve at least one of the boson propagators spanned over more than two external $\mu$-legs.

In 1D, after taking into account these leading contributions, the result  is
\begin{multline}
\Omega^{\text{1D}}_{\text{NA}}=\frac{\mu^2}{4\pi v_f}\left(1+\frac{\lambda_{bf}^2}{2\pi m^2v_f}\right)\ln{\frac{|\mu|}{v_f^2}}\\
+\frac{\lambda_{bf}^2\mu^2}{4\pi^2 m^2v_f^2}\left(\ln\frac{|\mu|}{v_f\Lambda_c}\right)^2\\
+\text {higher order terms}
\end{multline}
which is valid when $(\lambda^2 \ln |\mu|)/m^2$ is small.
In the hypothetical limit where $\mu$ becomes exponentially small so that $(\lambda^2_{bf} \ln |\mu|)/m^2$ is no longer small, one then has to further take into account the running of the coupling constant $\lambda^2_{bf}$ due to higher order effects.  
It remains an open question how the marginal operators change the free Majorana fermion fixed point physics. However, we speculate that the order of phase transition remains the same, although the marginal divergent behavior can be very different.

\begin{figure}
\includegraphics[width=\columnwidth]{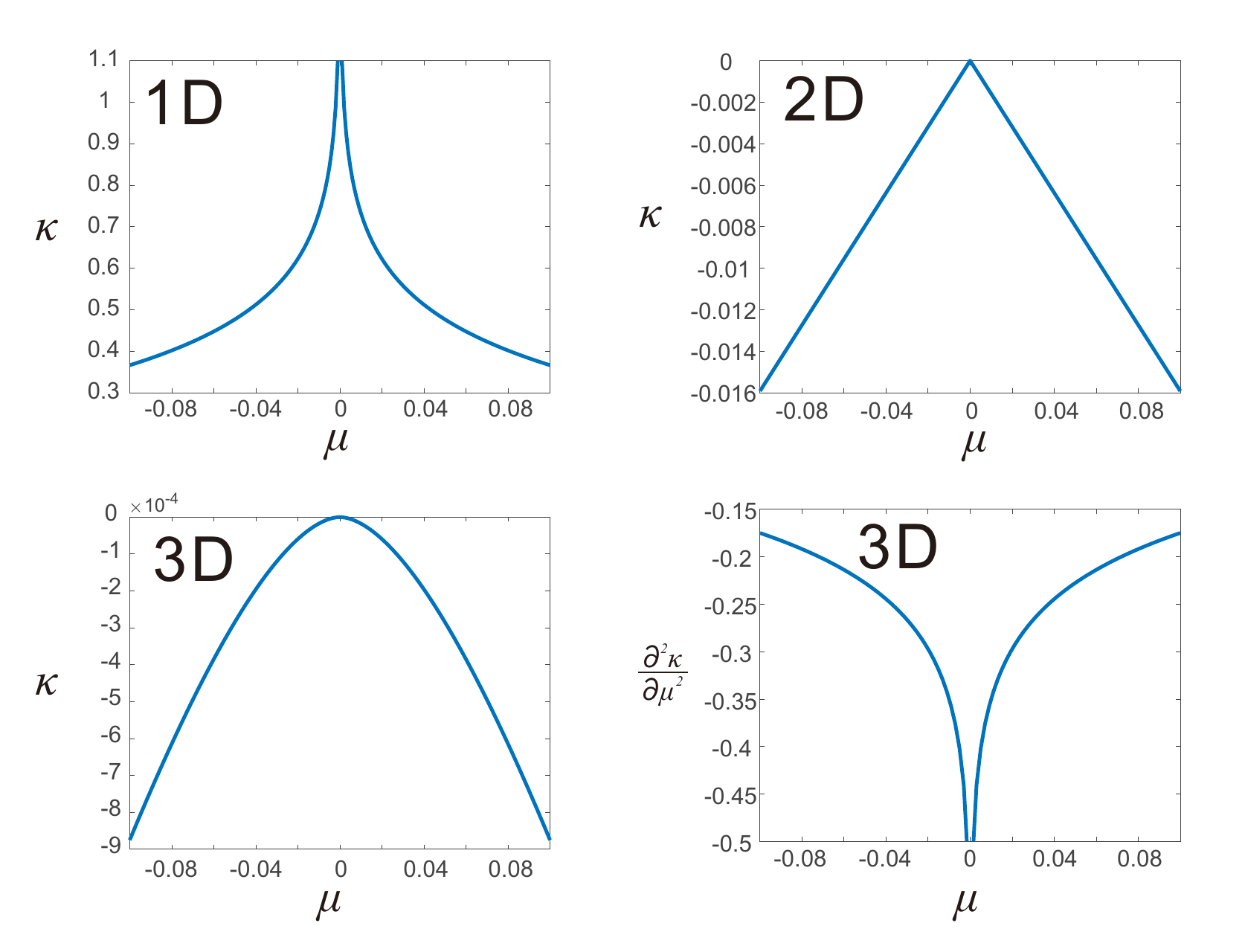}
\caption{Leading order non-analytical behaviours of the effective compressibility in different spatial dimensions. The compressibility $\kappa$ develops a cusp at $\mu=0$ where the topological phase transition occurs. In 1D, $\kappa$ is logarithmically divergent. In 2D and 3D, the compressibility is continuous but not analytical. To demonstrate the non-analyticity of $\kappa$ in 3D, we plot the second derivative of $\kappa$ with respect to $\mu$, which is divergent at $\mu=0$.\label{compressibility}}
\end{figure}

In conclusion, in $d$ dimensions, the topological phase transitions in isotropic superfluids involving Majorana fermions are $(d+1)$th order with corresponding cusps in thermodynamic potentials and the effective compressibility (see Fig. \ref{compressibility} for illustration). These structures are robust against interactions or fluctuations related to Majorana fermions \cite{1D}.

More cautions need to be made in the 1D effective field theory as the continuous symmetry cannot be broken spontaneously in one dimension and the arguments presented in Section III are not directly applicable.  
Nevertheless, the effective field theory can still be employed to characterize the IR physics and topological superfluid phase transitions in a wide range of lattice pairing models such as the Kitaev model where the symmetry is broken explicitly \cite{Kitaev01}. 
In the typical Kitaev nearest-neighbour pairing model and its extensions, the gauge symmetry is broken as an input by construction and serves as a starting point for considerations (i.e. disregarding one dimensional 
nature of phase dynamics). For those transitions that are effectively described by Majorana fermions, our analysis above suggests detailed structures of cusps and how interactions beyond the scope of those models can further affect the singularities.

However, since the traditional effective field theory is constructed out of local field operators, we do not anticipate the universality class of the effective model describe the transitions in various  long-range lattice pairing models \cite{Ortiz14,Pientka13,DeGottardi13,Viyuela18}. 
For instance, the order of transition here indeed differs from that in a 1D long-range pairing lattice model studied in Ref. \cite{Ortiz14}.
Moreover, the Majorana zero modes in our model have the standard exponential decay in real space instead of the algebraical decay structure suggested in the 1D long-range models \cite{Pientka13,DeGottardi13}.
The effective field theory for long-range pairing lattice models are likely to be constructed out of non-local field operators and remains to be studied in the future. 
In this article, we have restricted ourselves to superfluids and transitions with local fields only.
In recent years, there have also been efforts to find alternative ways to observe topological phase transitions that do not rely on the direct observation of Majorana boundary modes with an emphasis on the non-interacting 1D Kitaev lattice model or nano wires\cite{Kempkes16,Herviou17,Serina18}. Here we have demonstrated that the cusp structures developed in isotropic superfluids are generic consequences of the Majorana fermion universality class and they appear to be very robust against interactions and/or fluctuations.

\section{Some implications of cusps}
 
Majorana fermions have been puzzling particles that theorists and experimentalists have been searching for, for many years.
Almost massless neutrinos are a potential candidate for Majorana particles and remain to be verified or disproved to be true Majorana particles existing in our universe.
In the context of contemporary condensed matter physics, the exploration of Majorana physics in various crystals as an emergent phenomenon has been mostly focused on the observation of 
zero energy Majorana modes bound or attached to an edge or an interface. Tremendous efforts have been made to detect these modes and impressive progress has already been achieved.

Having freely propagating gapless Majorana fields/fermions in 2D and especially 3D systems, not particularly bound to a physical edge or surface, is appealing, but can only be realized in very limited physical systems.
Detecting and visualizing these exotic emergent particles in solids can be very exciting opportunities. The cusp structures near topological phase transitions obtained in this article are, in general, uniquely associated with both the existence of gapless Majorana fermions at the critical points, and more importantly, the tunability of Majorana masses via varying chemical potentials. 
The chemical potential can be physically controlled by electric gates in solid state systems, or for a system of fixed densities by tuning interactions or fermion density. Therefore, to a certain extent one can consider the cusps as one of the alternative approaches for detecting the emergent Majorana fermions and especially the variability of Majorana fermion masses in the emergent many-body phenomena. 
In contrast, in Dirac (or Weyl) semi-metals,  varying the chemical potential does not lead to a change of Dirac masses, at least not in a simple non-interacting picture. This is distinct from the response of Majorana masses to chemical potentials. For instance, in 3D topological Dirac semimetals, we do not expect a cusp in the derivative of the compressibility near half-filling.

Another interesting issue is whether the effective theory is applicable in physical one dimensional superfluids where the continuous symmetry cannot be broken in the same way as in high dimensions.
To avoid this dynamic issue, a possibility is to embed the 1D dynamics in a 2D geometry so that pairing order parameters effectively live in 2D to suppress strong fluctuations. Alternatively, one can also have a long but finite 1D system to suppress the long wavelength fluctuations. In this case, the effective theory shall correctly describe the low energy physics in the mesoscopic structure, subject to a finite size effect which will partially smear out the cusp structure near the critical point. We anticipate that the effective theory can still be extended to describe those transitions if the underlying physics is local in nature.

\section{conclusion}

In conclusion, we have identified and classified cusps near phase transitions between topologically different isotropic superfluids.
These transitions are all characterized by massless Majorana fermions interacting with fluctuation fields in different spatial dimensions.
We argue these are alternative approaches to detecting free propagating Majorana fermions in the bulk with emergent Majorana masses sensitive to many-body physics.
In recent years, there have been interesting attempts to establish duality connections between fermionic and bosonic representations in 2D \cite{Senthil04a,Senthil04b,Seiberg16,Metilitski17,Wang17}. It remains to be fully understood how to approach these phase transitions in  the framework of dual theories. In addition, as mentioned above although these transitions are generically non-supersymmetric, 
it is intriguing to ask whether the transitions can be physically tuned to be supersymmetric and under what energetic constraints this can happen\cite{Fendly03,Lee07,Yue10,Grover14,Jian15}. 
On the other hand, the edge states and surface states in topological superfluids (i.e. with $\mu <0$) are robust and massless. Thus, they can be ideal candidates for realizing supersymmetric conformal fields if further interacting 
with bosonic fluctuation fields. This possibility of applying topological boundary states to create supersymmetric fields was suggested before in Ref. \cite{Grover14}. We plan to explore these issues in the future in other concrete many-body systems.

\begin{acknowledgments}
One of the authors (F. Z.) would like to thank Ian Affleck, Leon Balents, Anton Burkov, Hae-Young Kee, Steven Kivelson, and Gordon W. Semenoff for helpful discussions. F. Z. would like to thank Leon Balents for pointing out the relations to dual theories.
F. Z. is a fellow of Canadian Institute for Advanced Research.
\end{acknowledgments}

\appendix

\section{Resummation and non-analyticities in grand potential}\label{resummation}

We compute the non-analytical terms in the grand potential from the fixed point action.
We sum up all one-loop diagrams in different dimensions to obtain the general form of the grand potential (in Euclidean space). 

For 2D $p_x\pm ip_y$ chiral superfluids,
\begin{align}
\Omega^\text{2D}&=\frac12\int\frac{d\omega}{2\pi}\frac{d^2k}{(2\pi)^2}\sum_{n=1}^\infty \frac1{2n}\text{Tr}\left[\left(\mu G_F(\omega,k)\tau_z\right)^{2n}\right]\nonumber\\
&=-\frac12\int\frac{d\omega}{2\pi}\frac{d^2k}{(2\pi)^2}\ln\left(1+\frac{\mu^2}{\omega^2+v_f^2k^2}\right);
\end{align}
where $G_F$ is the Green's function for spinless Nambu fermions in 2D. The integration yields
\begin{equation}
\Omega^\text{2D}=\frac{|\mu|^3}{12\pi v_f^2}+\text{analytical terms}.
\end{equation}
For 2D TRS superfluids, one needs to take the spin degree of freedom into consideration, and the grand potential is
\begin{align}
\Omega^\text{2D}_\text{TRS}&=\frac12\int\frac{d\omega}{2\pi}\frac{d^2k}{(2\pi)^2}\sum_{n=1}^\infty \frac1{2n}\text{Tr}\left[\left(\mu G_F^\text{(N)}(\omega,k)\tau_z\otimes I\right)^{2n}\right]\nonumber\\
&=\frac{|\mu|^3}{6\pi v_f^2}+\text{analytical terms},
\end{align}
where $G_F^\text{(N)}$ is the Green's function for spinful Nambu fermions in 2D.

For 3D superfluids,
\begin{align}
\Omega^\text{3D}&=\frac12\int\frac{d\omega}{2\pi}\frac{d^3k}{(2\pi)^3}\sum_{n=1}^\infty \frac1{2n}\text{Tr}\left[(\mu G_F^\text{(M3)}(\omega,k)\Gamma_0)^{2n}\right]\nonumber\\
&=-\frac{\mu^4}{16\pi^2 v_f^3}\ln\frac{|\mu|}{v_f^2}+\text{analytical terms},
\end{align}
where $G_F^\text{(M3)}$ is the Green's function for spinful Majorana fermions in 3D.

Lastly, for 1D spinless $p$-wave superfluids
\begin{align}
\Omega^\text{1D}&=\frac12\int\frac{d\omega}{2\pi}\frac{dk}{2\pi}\sum_{n=1}^\infty \frac1{2n}\text{Tr}\left[\left(\mu G_F^\text{(M1)}(\omega,k)\tau_y\right)^{2n}\right]\nonumber\\
&=\frac{\mu^2}{4\pi v_f}\ln\frac{|\mu|}{v_f^2}+\text{analytical terms},
\end{align}
where $G_F^\text{(M1)}$ is the Green's function for spinless Majorana fermions in 1D.

Here, we have omitted the explicit form of analytical terms in all cases, since their exact form also depends on the UV physics beyond the scope of the effective theory. In contrast, the leading order non-analytical term is dictated only by the infrared physics described by the fixed point Lagrangian.


\begin{thebibliography}{References}
\bibitem{mu} The effective chemical potential can differ from the physical chemical potential by a meanfield shift. This difference plays little role in our analyses.
\bibitem{1D}  Although in 1D, we have found the interactions can be marginally relevant in exponentially small energies.
\end{thebibliography}
\end{document}